\documentclass{jaa}
\usepackage{natbib}
\usepackage{graphicx}
\usepackage[para,online,flushleft]{threeparttable}
\usepackage{subfigure}
\usepackage{amsmath}
\usepackage{amssymb}
\usepackage{multirow}
\usepackage[mathscr]{euscript}
\usepackage[para,online,flushleft]{threeparttable}
\usepackage{multirow}
\usepackage{booktabs}
\usepackage{xr}
\usepackage{array}
\usepackage{graphicx}
\usepackage[dvipsnames]{xcolor}
\usepackage{multicol}
\usepackage{colortbl}
\usepackage{xcolor}

\newcolumntype{L}[1]{>{\raggedleft\arraybackslash}m{#1}}

\begin{document}\sloppy

\title{Modeling of thermal and non-thermal radio emission from HH80-81 jet}


\author{Sreelekshmi Mohan\textsuperscript{*}, S. Vig \textsuperscript{} and S. Mandal\textsuperscript{}}
\affilOne{\textsuperscript{}Indian Institute of Space Science and Technology, Thiruvananthapuram, 695 547, India}


\twocolumn[{

\maketitle

\corres{sreelekshmimohan94@gmail.com}

\msinfo{1 January 20xx}{1 January 20xx}

\begin{abstract}
Protostellar jets are one of the primary signposts of star formation. A handful of protostellar objects exhibit radio emission from ionized jets, of which a few display negative spectral indices, indicating the presence of synchrotron emission. In this study, we characterize the radio spectra of HH80-81 jet with the help of a numerical model that we have developed earlier, which takes into account both thermal free-free and non-thermal synchrotron emission mechanisms. For modeling the HH80-81 jet, we consider jet emission towards the central region close to the driving source along with two Herbig-Haro objects, HH80 and HH81. We have obtained the best-fit parameters for each of these sources by fitting the model to radio observational data corresponding to two frequency windows taken across two epochs. Considering an electron number density in the range $10^3 - 10^5$~cm$^{-3}$, we obtained the thickness of the jet edges and fraction of relativistic electrons that contribute to non-thermal emission in the range $0.01^{\circ} - 0.1^{\circ}$ and $10^{-7} - 10^{-4}$, respectively. For the best-fit parameter sets, the model spectral indices lie in the range of -0.15 to +0.11 within the observed frequency windows.
\end{abstract}


}]


\doinum{12.3456/s78910-011-012-3}
\artcitid{\#\#\#\#}
\volnum{000}
\year{0000}
\pgrange{1--}
\setcounter{page}{1}
\lp{1}

\section{Introduction}
Jets are launched along the axis of a rotating system as the outcome of accretion of materials from the surrounding interstellar medium (ISM). Accretion is aided by jets, by acquiring the excess angular momentum \citep{1982MNRAS.199..883B,1983ApJ...274..677P} and carrying it away. These jets could range from ultrarelativistic/relativistic jets in high energy astrophysical phenomena \citep{doi:10.1146/annurev.astro.37.1.409,doi:10.1146/annurev-astro-081817-051948} to non-relativistic jets in protostars \citep{Reipurth_2004,Lee_2007} and brown dwarfs \citep{2009ASSP...13..259W,10.1093/mnras/stu1461}. Observational as well as modeling studies have shown that protostellar jets can either promote or hamper star formation activity in their immediate neighbourhood \citep{2006ApJ...640L.187L, 2007ApJ...662..395N,2008ApJ...683..255S} and also the surrounding ISM through the transfer of momentum and energy \citep{2006A&A...453..911F}.

  Jets are considered to be closely related to the structure of the magnetic field within the protostellar system. Upon launch, the jet collimation is achieved by the limiting pressure of toroidal component of helical magnetic fields \citep{1997ASPC..121..845L,Meier84} which has been confirmed by various hydromagnetic and magnetohydrodynamic models \citep{1987ApJ...315..504L,1995ApJ...439L..39U, Cerqueira_2001,zanni2004mhd,Bellan2005}. Theories predict similar frameworks for the jet magnetic fields, regardless of them being of protostellar or AGN origin \citep{1986NYASA.470...88K,2000AIPC..522..275L}. The direction and strength of the magnetic field can be determined with the knowledge of synchrotron emission parameters \citep{2010Sci...330.1209C}. Low noise sensitive observations towards a massive protostellar jet had revealed the presence of linearly polarized emission, which confirmed the underlying synchrotron emission mechanism \citep{2010Sci...330.1209C}. The synchrotron emission is prominent along the edges and jet termination points, where the jet interacts against the ambient medium resulting in strong shocks, thereby facilitating particle acceleration \citep{2017ApJ...851...16R}.

The theory of synchrotron radiation from high-energy phenomena associated with extragalactic jets is well-established \citep{1982ApJ...256...13R,1987ApJ...322..643B,1998ApJ...497L..17S} owing to the abundance of available observational evidence \citep{1999ApJ...523..177W,2000ApJ...543..373D,2006ARA&A..44..463H}. However, the origin of synchrotron emission from protostellar jets is still not clear. It is possible that the lower energies/velocities of these jets compared to that of relativistic jets could result in the sparse detection of synchrotron emission from them \citep{2003MNRAS.339.1223P}. 

Herbig-Haro objects are the outcome of the impact of protostellar jets on the ISM, which manifests in the form of a chain of well-defined knots or lobes. Our understanding regarding the formation of knots is still not complete, although models have considered episodic accretion and/or variation in ejection velocities to explain their presence \citep{{2004ApJ...606..483L},{2010A&A...511A..42B}}. Although the former model can generate a train of knots, it is unable to explain the velocity structure and continuous nature of the jet near the driving source as identified by observations \citep{2016ApJ...816...32J}. Therefore, the latter model is the widely accepted one, despite the fact that the mechanisms responsible for variation in jet velocity are currently poorly understood. Observations indicate the presence of non-thermal emission from knots of a few YSO jets, while in most cases, the emission can be explained using thermal free-free mechanism. For thermal jets, the most widely used model is that of \citet[][hereafter Reynolds model]{1986ApJ...304..713R} model, which analytically calculates radio emission and spectral indices using thermal free-free emission for different jet geometries. On the other hand, the numerical model that we had developed earlier \citep{2022MNRAS.514.3709M} incorporates a more general geometry and calculations compared to the Reynolds model. In this paper, we apply the model to the radio spectrum of HH80-81 jet, which is known to display negative spectral indices, to characterize the properties of the jet material.

\section{The model} \label{model}

In this section, we briefly discuss the model and model parameters. Our model \citep{2022MNRAS.514.3709M} offers more generalizations as compared to the Reynolds model in characterizing the radio spectra of protostellar jets. The Reynolds model is only applicable for narrow collimated jets, and it calculates flux densities such that the base of the jet is fully optically thick and the farther regions are fully optically thin, the sum of which contributes to the total flux density of the jet. On the other hand, our model considers intermediate optical depth values, has more flexible geometry and is applicable to narrow as well as wide opening angle jets. The introduction of non-thermal synchrotron radiation is the major addition in our model.

The application of our model can be two-fold. It can be utilized to model radio emission from knots observed farther away from young stellar objects (YSOs) as well as the thermal jets at closer radial distances. However, we would like to draw attention to the fact that our model does not incorporate the contribution of dust emission, which could be significant for radio frequencies $\gtrsim50$~GHz.

\subsection{Jet geometry and model parameters}

\begin{figure}
	\includegraphics[width=\columnwidth]{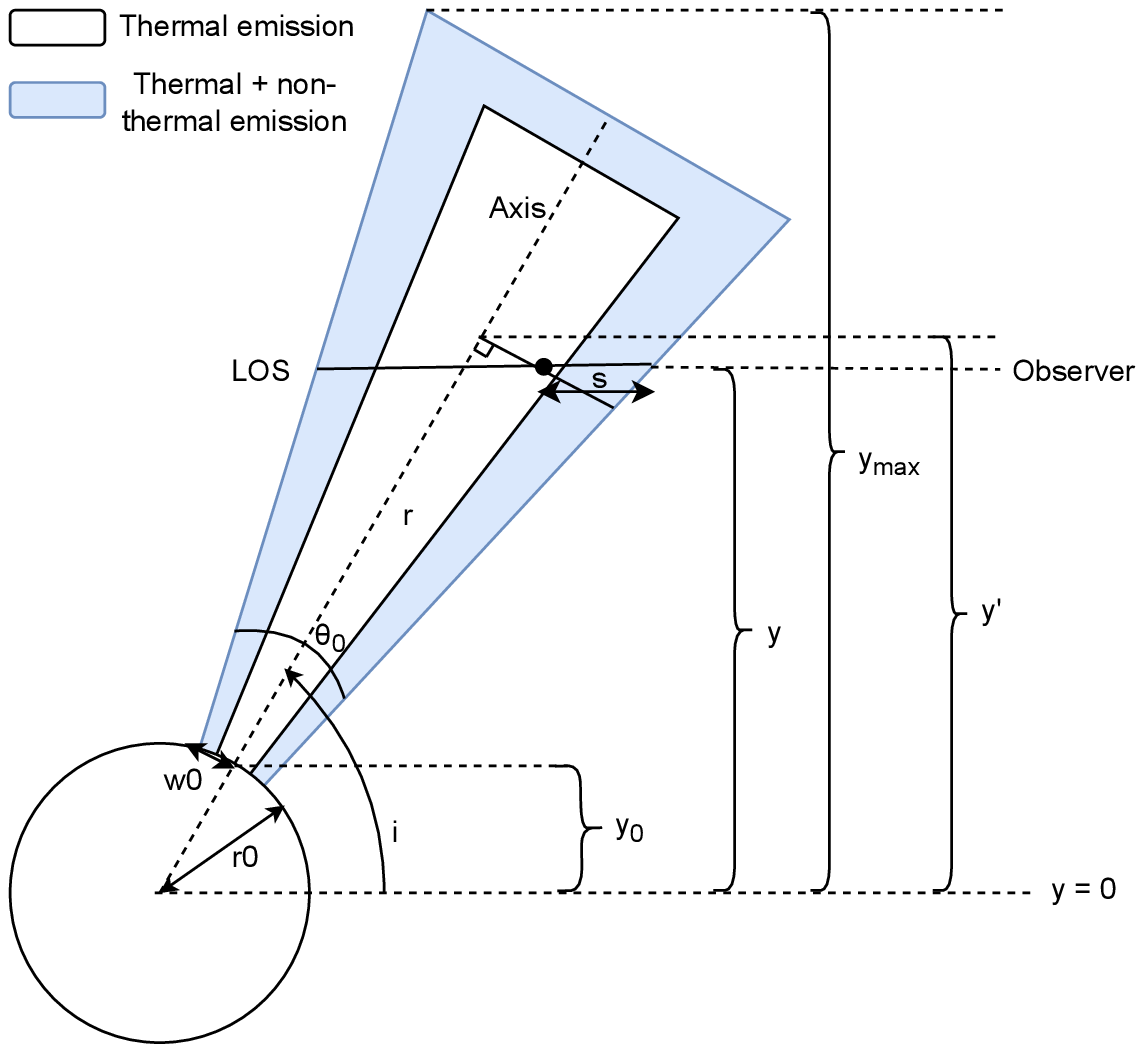}
    \caption{Schematic diagram of a jet with constant opening angle, adopted from \citet{2022MNRAS.514.3709M}. The blue-shaded area represents the shocked region of the jet material that contributes to a combination of thermal and non-thermal emission. The unshaded jet region closer to the axis represents the fully thermal jet. The variables used in the model are also shown here.}
    \label{fig:geom}
\end{figure}

For the jet parameters, we use the same terminology as that of \citet{2022MNRAS.514.3709M}. Fig.~\ref{fig:geom} depicts the jet's geometry and orientation with respect to the observer. The distance from the central source at which the jet is sufficiently ionized for free-free emission to be observable is represented by $r_0$. We presumptively inject the jet at $r_0$ with an opening angle of $\theta_{0}$ and a half-width of $w_{0}$ at $r_0$. The projected distance corresponding to $r_0$ and the total length of the jet are given as $y_0$ and $y_{max}$, respectively. The thin outer shell (the blue-shaded area in Fig.~\ref{fig:geom}) represents the material that emits both thermal and non-thermal synchrotron emissions. We represent the inclination angle of the jet with respect to the observer's direction as $i$. The following relationships determine the length of the jet projected in the plane of the sky, $y$, and the jet's half-width ($w$), at any radial distance $r$. 

\begin{equation}\label{eq:y}
y = r\, \sin{i}
\end{equation}
\begin{equation}\label{eq:width}
w(y) = w_{0} \bigg(\frac{r}{r_{0}}\bigg)^{\epsilon} = w_{0} \bigg(\frac{y}{y_{0}}\bigg)^{\epsilon}
\end{equation}
 
Here, $\epsilon$ is the radial power-law index for the width profile. We first present the equations associated with the free-free emission model. The emission and absorption coefficients associated with free-free emission are given by, 
\begin{equation}\label{ff_jnu}
j^{ff}_\nu\, = a_j\, n^2\, x^2\, T^{-0.35}\, \nu^{-0.1}
\end{equation}
\begin{equation}\label{ff_alphanu}
\alpha^{ff}_\nu\,=a_k\, n^2\, x^2\, T^{-1.35}\, \nu^{-2.1}
\end{equation}

Here, $n$, $x$ and $T$ are the electron number density in the jet material, ionization fraction and electron temperature, each having a power-law profile in the radial direction with power-law indices given by  $q_n$, $q_x$ and $q_T$, respectively. The proportionality constants of emission and absorption coefficients, expressed in cgs units, are $a_j\,=6.50 \times 10^{-38}\,$~ergs~cm$^{3}$~Hz$^{-0.9}$~K$^{0.35}$~s$^{-1}$~sr$^{-1}$ and $a_k\,=0.212$~cm$^{5}$~K$^{1.35}$~Hz$^{2.1}$. For a source function given by $\frac{j^{ff}_{\nu}(s)}{\alpha^{ff}_{\nu}(s)}$, the flux density along a line-of-sight (LOS) of the jet that subtends an incremental solid angle d$\Omega$ with respect to the observer is given by,
\begin{equation}\label{eq:fluxGeneral}
S_{\nu} = \int_{y_{0}}^{y_{max}} d\Omega\, \int_{0}^{\tau_{\nu}} \frac{j^{ff}_{\nu}(s)}{\alpha^{ff}_{\nu}(s)}\, (e^{-(\tau^{ff}_{\nu} - \tau (s)) })\, d\tau
\end{equation}

Here, $s$ represents the arbitrary location of any point along a LOS. $\tau_{\nu}^{ff}\, = \, \int \alpha_{\nu}(s) \, ds$ and $\tau(s)$ are the total LOS optical depth and the optical depth corresponding to any $s$ along a LOS, respectively.

We incorporate synchrotron emission in the model in a similar manner as free-free emission. For an arbitrary Lorentz factor $\gamma$, the number density of non-thermal electrons $n(\gamma)d\gamma$ between $\gamma$ and $\gamma+d\gamma$ is given below.
\begin{equation}\label{eq:num_dist}
n(\gamma)\,d\gamma = n_{k}\,\gamma^{-p}\,d\gamma
\end{equation}   
   
Here, $p$ is the number density distribution index of the non-thermal electron population. For a magnetic field strength $B$, and a distribution of electrons with mass $m$, charge $e$, the synchrotron emission coefficient, $j^{syn}_{\nu}$, is given by the following expression \citep{rybicki2008radiative}.

\begin{equation} \label{eq:emission}
\begin{split}
j^{syn}_{\nu}& = \frac{1}{4\pi} \frac{\sqrt{3}\, e^{3}\, n_k\, B\, \sin{\alpha_\text{pa}} }{2\pi\, m\, c^{2}\, (1+p)}\, \Gamma\left(\frac{p}{4} + \frac{19}{12}\right)\, \Gamma\left(\frac{p}{4} - \frac{1}{12}\right) \times\\
 &  \bigg(\frac{2\pi\, \nu\, m\, c}{3e\, B\, \sin{\alpha_\text{pa}}}\bigg)^{-(p-1)/2} 
\end{split}
\end{equation}

Here, $n_k$ is the proportionality constant for the number density of electrons, $\alpha_\text{pa}$ is the angle between the magnetic field and the electron velocity, and $\Gamma(x)$ is the gamma function for given $x$ and $\alpha_\text{pa}$ parameters. Similarly, the synchrotron absorption coefficient, $\alpha^{syn}_{\nu}$, is given as,

\begin{flushleft}
\begin{equation} \label{eq:absorption}
\begin{split}
\alpha^{syn}_{\nu} &= \frac{\sqrt{3} e^{3}}{8\pi\, m}\, \bigg(\frac{3e}{2\pi m^{3} c^{5}}\bigg)^{p/2}\, n_k\, (B\, \sin{\alpha_\text{pa}})^{(p+2)/2}\, \Gamma \left(\frac{3p+2}{12}\right) \\
 & \times \Gamma \left(\frac{3p+22}{12}\right)\, \nu^{-(p+4)/2}
\end{split}
\end{equation}
\end{flushleft}

In the model, we combine the free-free and synchrotron emission contributions along the extremities of the jet in order to determine the radio flux densities produced by the shocked jet material. The following expressions represent the combined emission and absorption coefficients in the shocked region. 

\begin{equation}\label{eq:jcoeff_coupled}
j^{ff+syn}_{\nu}(s,y) = j^{ff}_{\nu}(s,y) + j^{syn}_{\nu}(s,y)
\end{equation}
\begin{equation}\label{eq:alcoeff_coupled}
\alpha^{ff+syn}_{\nu}(s,y) = \alpha^{ff}_{\nu}(s,y) + \alpha^{syn}_{\nu}(s,y) 
\end{equation}

\par The model also incorporates an ionization fraction profile that decreases across the width of the jet. For any projected length $y^{\prime}$ (Fig.~\ref{fig:geom}), which is a function of $s$, the variation of ionization fraction across the given LOS of the jet can be introduced through a power-law with index $q_x'$ as follows.
\begin{equation}\label{eq:x}
x(s) = x_a(y')\left[\frac{w(y')}{w(y') - w'(y')}\right]^{q_x'}
\end{equation}

\noindent where $x_a(y')$ is given by,
\begin{equation}\label{eq:x_axial}
x_a(y') = x_0 \left(\frac{y'}{y_0}\right)^{q_x} 
\end{equation}
\noindent Here, the subscript $a$ implies that it is axial, $x_0$ represents the ionization fraction at $r_0$ and $w^{\prime}(y^{\prime})$ represents the perpendicular distance of any point $s$, from the jet axis.

\section{Comparison of model with observations} \label{apply_ysos}

\begin{figure}[!h]
	\hspace{-0.1cm}
	\includegraphics[scale=0.5]{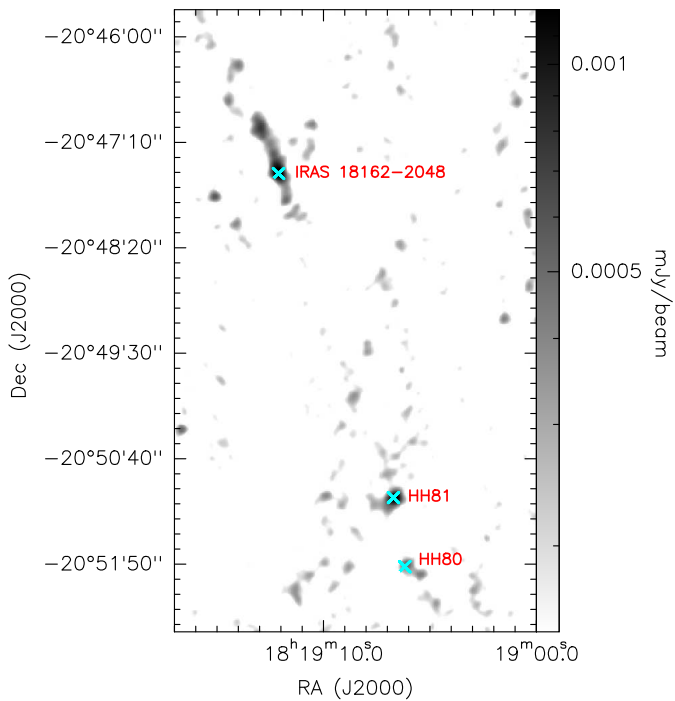}
    \caption{610~MHz radio continuum image of IRAS~18162-2048 region showing the HH80-81 jet, adopted from \citet{2018MNRAS.474.3808V}. The central region of the jet (IRAS~18162-2048) and Herbig-Haro objects HH80 and HH81, which are investigated in this study, are marked in cyan crosses.}
    \label{fig:vig2018}
\end{figure}

In this section, we employ our model to understand the radio spectra of a prominent protostellar jet, HH80-81, which exhibits the characteristics of non-thermal emission. The HH80-81 protostellar jet is the largest known and most luminous Herbig-Haro system in the Galaxy. It is powered by the protostar associated with the source IRAS~18162-2048, the radio emission of which indicates that IRAS~18162-2048 is a B0 type protostar located at a distance of $1.4$~kpc \citep{2020ApJ...888...41A}. The HH80-81 jet is highly collimated with an approximate size of $18.7$~pc \citep{2012ApJ...758L..10M,2018MNRAS.474.3808V}. The inclination angle of the jet with respect to the observer's direction was observed to be  $34^\circ$ \citep{1998AJ....116.1940H}. The jet consists of numerous knots, of which the most prominent ones include HH80, HH81 towards the south and HH80N in the north \citep{1989RMxAA..17...59R,2001ApJ...562L..91G}. Radio and optical measurements have indicated velocities of $600-1000$~km/s for these knots \citep{1995ApJ...449..184M,1998AJ....116.1940H}. The radio continuum image of the HH80-81 jet at 610~MHz, originally published by \citet{2018MNRAS.474.3808V}, is presented in Fig.~\ref{fig:vig2018}. Here, we have only shown the region of interest for our investigation comprising of the central jet along with the HH80 and HH80 knots, which are marked with cyan crosses and labeled accordingly in the figure.

In the immediate vicinity of the exciting source, thermal emission is observed, and synchrotron emission has been confirmed farther out up to a distance of $\sim0.5$~pc through the detection of linearly polarized emission \citep{2010Sci...330.1209C}. The outer lobes along the jet have also been observed to show non-thermal nature. Non-thermal emission at lower frequencies of 325, 610 and 1300~MHz was detected by \citet{2018MNRAS.474.3808V}, which is the first case of detection of non-thermal emission from this jet at such low frequencies. In addition, variability of non-thermal emission was also demonstrated by them. The measured spectral indices of most of the knots are as steep as $\sim-0.7$. Both the polarization observations \citep{2010Sci...330.1209C} and non-thermal spectral indices \citep{2018MNRAS.474.3808V} have provided estimates of the strength of magnetic field to be $\sim200$~$\mu$G.  

\begin{figure*}
\hspace*{3cm}
\minipage{\textwidth}
 \includegraphics[scale = 0.7]{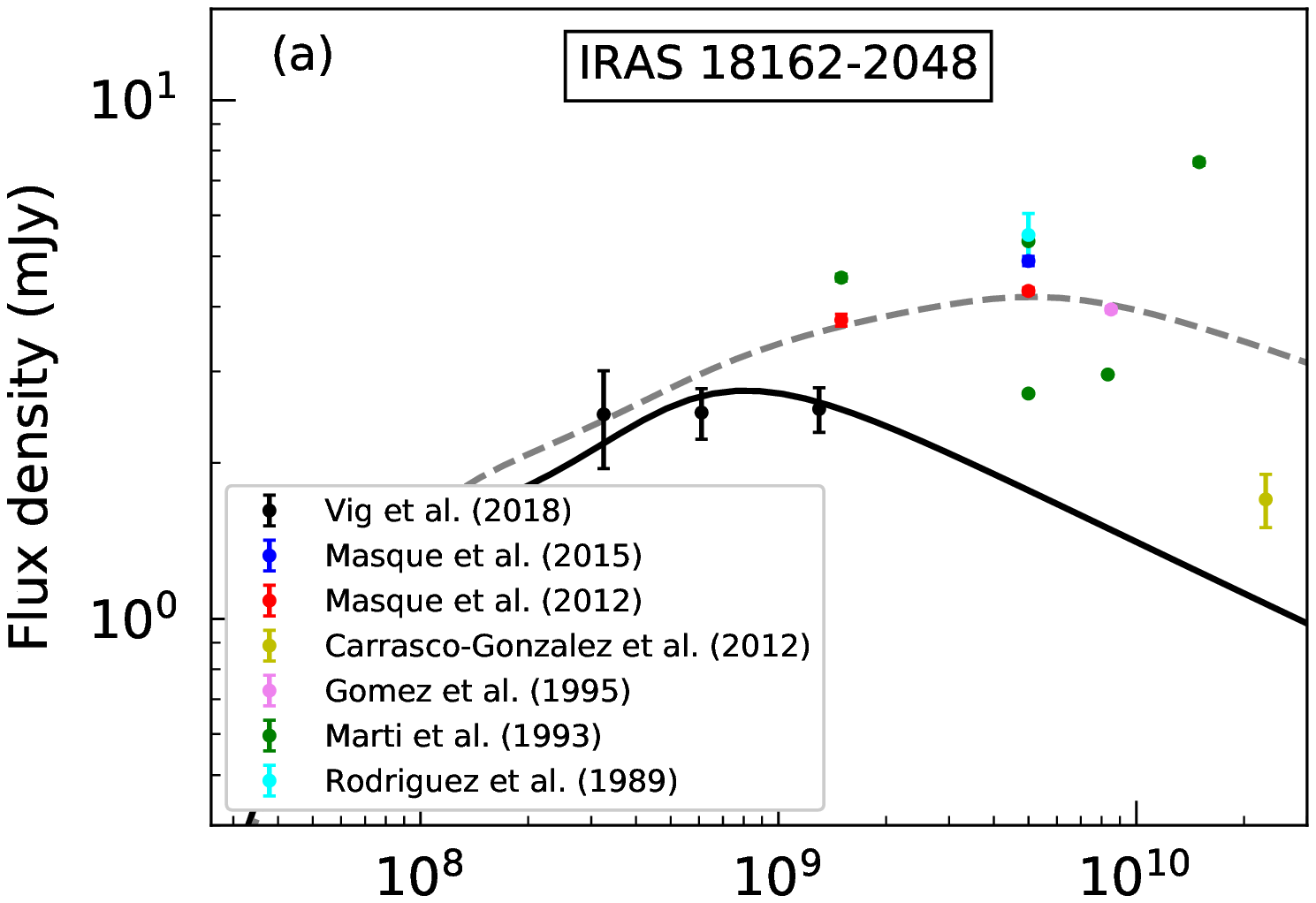}
\endminipage
\centering
\vskip +0.1cm
  \hspace*{3cm}
\minipage{\textwidth}
 \includegraphics[scale = 0.7]{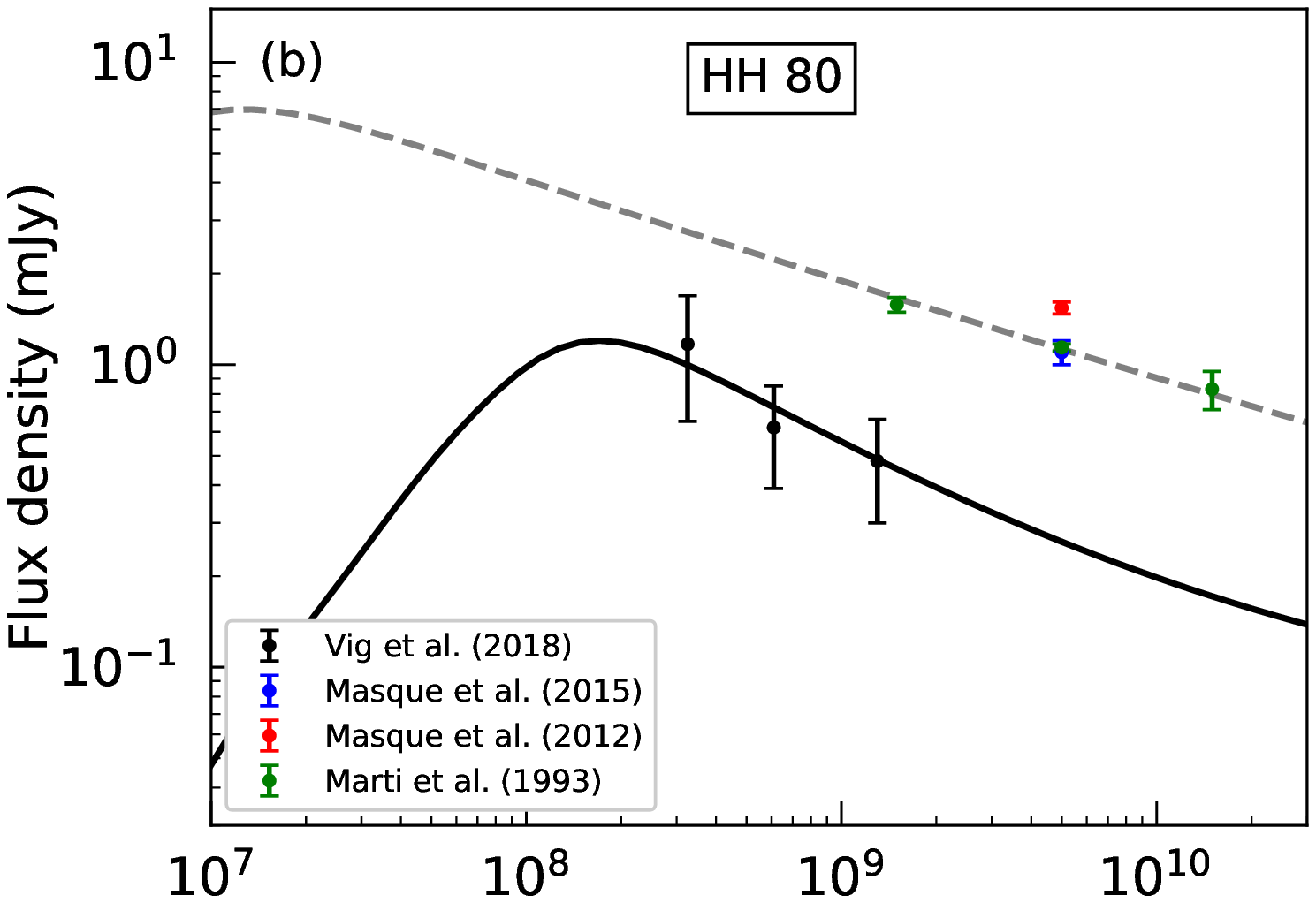}
\endminipage

\centering
\vskip +0.1cm
  \hspace*{3cm}
\minipage{\textwidth}
 \includegraphics[scale = 0.7]{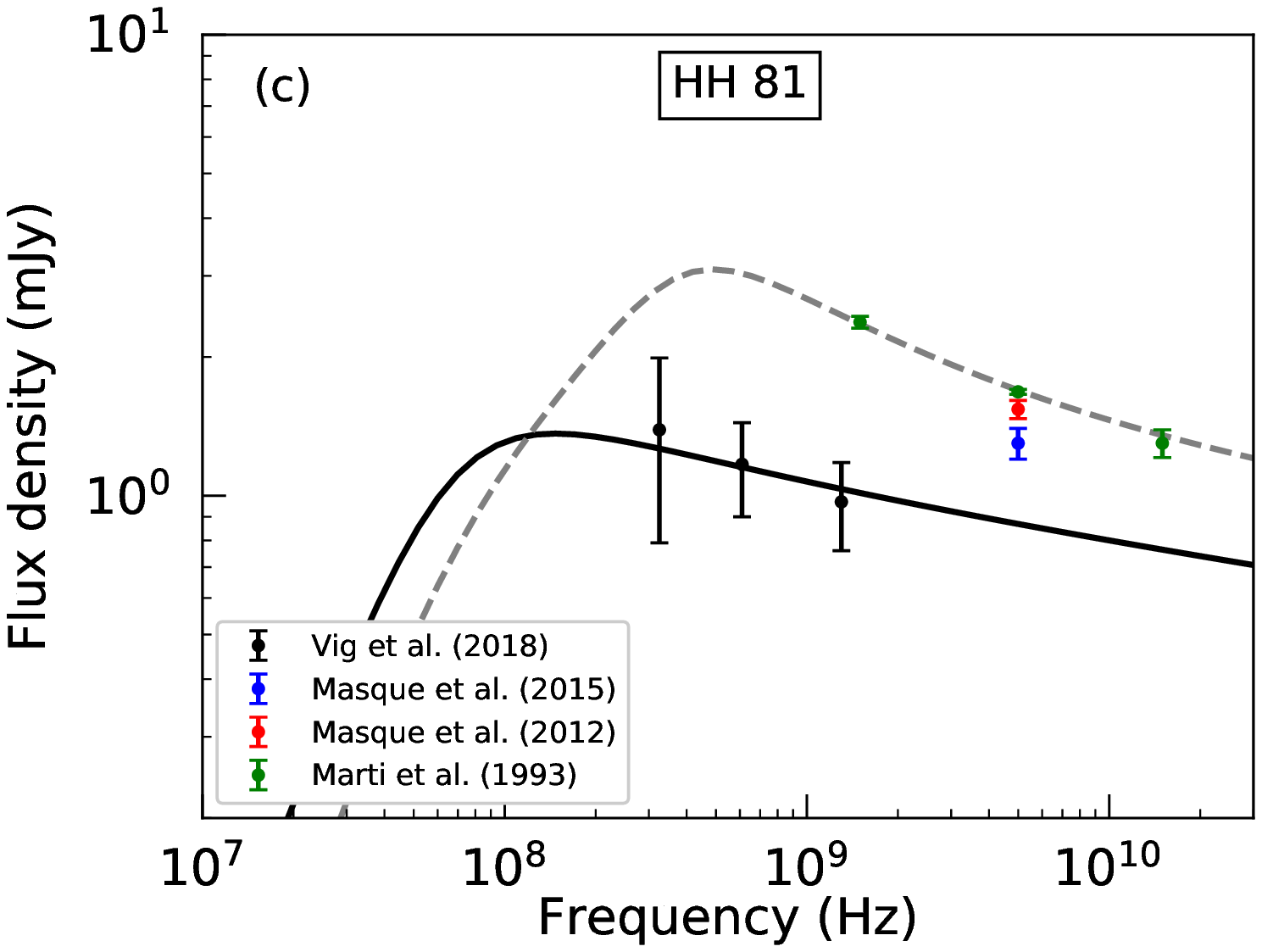}
\endminipage

\caption{Comparison of best fit-model spectra with observational data for (a) IRAS~18162-2048, (b) HH80, and (c) HH81. The data points shown in black and red are used for fitting, and the remaining data are plotted only for representation. The solid black curves in (a), (b) and (c) represent models 18162/2016, HH80/2016 and HH81/2016, respectively. The dashed grey curves in (a), (b) and (c) represent models 18162/2009, HH80/1989 and HH81/1989, respectively.}
\label{fig:model_spec}
\end{figure*}

As a case study, we choose to investigate three prominent lobes of this jet system: the central region IRAS~18162-2048, as well as the knots HH80 and HH81 using the radio measurements, where the flux densities have been measured at similar epochs. These sources are labeled by their respective names in Fig.~\ref{fig:vig2018}. The references for the observed radio flux densities of the knots are given in Fig.~\ref{fig:model_spec}. The sizes and locations of jet lobes, the jet opening angles, and other properties such as estimates of magnetic field, which are taken from previous studies in literature, where available, are listed in Table.~\ref{tab:table_modelParms_ref} along with the relevant references. These physical parameters are incorporated into the models to generate the radio spectra, and more details about the parameters are given in the subsequent subsections.

The physical parameters that are not constrained by the observations are decided based on a grid of values that are either generally observed in jets or expected from physical models of jets. These include $q_n$, $q_{x}$, $q_x'$, $\delta \theta$, $p$ and $\eta^{rel}_{e}$. As this is a highly collimated jet, we assume that the jet opening angle continues to decrease as it moves out radially and, therefore, for each knot, $\epsilon$ is kept constant. Also, since all the knots studied here belong to the same jet, we have assumed the same value of $\epsilon$ for all these knots. We also assume a temperature of $10^4$~K, typical of ionized gas, for all the knots. The parameters $n_0$ and $x_0$ are combined into a single parameter corresponding to the ionized number density ($n_0\times x_0$) and typical values as obtained from observations are used in the models. The rest of the parameters are explored in a range typical to jets through a coarse grid, and the best fitting parameter set is chosen through the chi-square minimization. A finer grid around this set was explored to choose the best-fitting model based on chi-square value. The set of values explored for each parameter in the coarse grid are listed in Table.~\ref{tab:table_grid}. As the observational data is limited and we are interested in obtaining typical model parameters, the grid step size can be considered as representative of the parameter uncertainties. The observational data for the selected knots and their corresponding best-fit models are shown in Fig.~\ref{fig:model_spec}, and each case is discussed below.


\begin{table*}
\caption{Physical parameters of the jet knots based on observations.}
\begin{center}
\begin{threeparttable}
\begin{tabular}{ l L{2cm} L{2cm} L{1cm} L{1cm} L{1cm} L{1cm} L{1cm} L{1cm} L{1cm} L{1cm} L{1cm} L{1cm} } \hline \hline	
\raggedright
Object &  $r_{0}^a$   &  $r_{max}^b$   &  $\theta_{0}^c$  &  $T_{0}^d$ &  $B_{0}^e$  & i$^f$ &  $d^g$ & References$^\dagger$\\
 &  (au) &  (au) & ($^\circ$)  & (K)  &    ($\mu$G)  & ($^\circ$) & (pc) &  &  \\ \hline 
IRAS~18162-2048 & 250 & 826 & 38.5 & $10^4$ &  200 & 34 & 1400 & 1,2,3,4,5,6,7 \\
HH80 & 657741 & 675672 & 2.4 &  $10^4$ &  132 &  34 &  1400 & 1,2,7,8,9 \\
HH81 & 553636 & 562004 & 2.5 &  $10^4$ &  160 & 34  & 1400 & 1,2,7,8,9 \\
\hline \
\end{tabular}
\label{tab:table_modelParms_ref}
\begin{tablenotes}

\tiny{
 $^a$ The projected distance of the knot from the central source that corresponds to the injection radius for continuous jet given in angular units \\  
 $^b$ The projected distance of the farther end of the knot\\ 
 $^c$ The jet opening angle at $r_0$ \\ 
 $^d$ The electron temperature of the jet material at $r_0$\\ 
 $^e$ Jet magnetic field at $r_0$\\
 $^f$ Inclination angle \\
 $^g$ distance of the source from the observer in pc\\
 $^\dagger$(1) \citet{2018MNRAS.474.3808V} (2) \citet{1993ApJ...416..208M} (3) \citet{2010Sci...330.1209C} (4) \citet{2012ApJ...752L..29C} (5) \citet{2012ApJ...758L..10M} (6) \citet{1989ApJ...346L..85R} (7) \citet{2020ApJ...888...41A} (8) \citet{2010A&A...511A...8B} (9) \citet{1998AJ....116.1940H}
 
 }
\end{tablenotes}
\end{threeparttable}
\end{center}
\end{table*}


\begin{table*}
\caption{Range of parameters explored in the grid of models.}
\begin{center}
\begin{threeparttable}
\begin{tabular}{l l } \hline \hline	
Parameters & Range Explored and Grid spacing\\ \hline 
Power-law index of radial number density variation ($q_n$) & -4 to -1 in steps of 0.5\\
Power-law index of radial ionization fraction variation ($q_x$) & -3 to 0 in steps of 0.5\\
Power-law index of lateral ionization fraction variation ($q'_x$) & -5 to 0 in steps of 0.5 \\
Angular thickness of shocked region ($\delta \theta$ [$^\circ$]) & 0.01 to 0.1 in steps of 0.01\\
Power-law index of non-thermal electron population ($p$) & 1.7 to 2.6 in steps of 0.1\\
Fraction of relativistic electrons ($\eta^{rel}_{e}$) & $10^{-7}$ to $10^{-4}$ by a factor of 10\\
\hline \

\end{tabular}
\label{tab:table_grid}
\end{threeparttable}
\end{center}
\end{table*} 

\begin{table*}
\caption{Best fit model parameters of the observed jets.}
\begin{center}
\begin{threeparttable}
\begin{tabular}{ l c c c c c c c c c c c} \hline \hline
Model & $\epsilon^a$ $^*$ & $n_0 x_0^b$ $^\dagger$  &  $q_n^{c}$ &  $q_x^d$ & $q'^{e}_x$ & $\delta \theta^f$ & $p^g$& $\eta^{rel}_{e}$ $^h$ & $\chi^{2}$ $^i$ & References$^\ddagger$\\
 &  &  & (cm$^{-3}$) &  &  &  & ($^{\circ}$) &  &   &   &     \\ \hline 
18162/2016  & 2/3 & 4.5$\times$10$^4$ &  -1 &  -0.2 & 0.0$^*$ & 0.10 & 1.7 & 1$\times10^{-4}$ & 0.87 & 1 \\
18162/2009  & 2/3 & 3$\times$10$^5$ &  -2 &  -0.5 & 0.0$^*$ & 0.10$^*$ & 1.7 & 1$\times10^{-4}$ & 0.60 & 2 \\
HH80/2016 & 2/3 & 200 & -2 & -2.5 & -3.5 & 0.01 & 2.3 &1$\times10^{-6}$  & 0.30 & 1\\
HH80/1989 & 2/3 & 200 & -3 & -0.2 & -3.0 & 0.01$^*$ & 1.7 & 1$\times10^{-6}$ &  1.07 & 3\\

HH81/2016 & 2/3 & $10^3$ & -3 &  -5.0 & -4.5 & 0.01 & 2.3 & 1$\times10^{-7}$ & 0.14 & 1\\
HH81/1989 & 2/3 & $10^3$ & -2 &  -0.2 & -4.0 & 0.01$^*$ & 2.4 &  1$\times10^{-6}$ & 0.96 & 3 \\

\hline \
\end{tabular}
\label{tab:table_modelParms_assumed}
\begin{tablenotes}
 \tiny{
 
 $^q$ Power-law index of radial variation in jet width \\
 $^b$ Ionized number density of the jet at $r_0$  \\  
 $^c$ Power-law index of radial variation in jet number density  \\ 
 $^d$ Power-law index of radial variation in ionization fraction \\
 $^e$ Power-law index of lateral variation in ionization fraction\\
 $^f$ The thickness of the outer shocked region in degrees \\
 $^g$ Power-law index of non-thermal electron distribution\\
 $^h$ Fraction of relativistic electrons in the shocked region\\
 $^i$ Chi-square value of the best-fit model\\
 $^*$ Assumed parameter \\
 $^\dagger$ Obtained from observations \\
 $^\ddagger$(1) \citet{2018MNRAS.474.3808V}  (2) \citet{2012ApJ...758L..10M} (3) \citet{1993ApJ...416..208M} 
  }
\end{tablenotes}
\end{threeparttable}
\end{center}
\end{table*}



\begin{table*}
\caption{Spectral features of the best-fit models of the observed jets.}
\begin{center}
\begin{threeparttable}
\begin{tabular}{l c c c} \hline \hline	
\raggedright
Source & Peak Flux (mJy)& Peak Frequency (MHz) & Spectral index$^*$\\ \hline 
IRAS~18162-2048$^1$ & 2.75 & 762 & +0.13 \\
IRAS~18162-2048$^2$ & 4.18 & 5311 & +0.11  \\
HH80$^1$ & 1.20 & 171 & -0.52  \\
HH80$^3$ & 6.97 & 13 & -0.32 \\
HH81$^1$ & 1.36 & 147 & -0.15  \\
HH81$^3$ & 3.10 & 486 & -0.24 \\
\hline \

\end{tabular}
\label{tab:specfeature}
\begin{tablenotes}
 \tiny{ (1) \citet{2018MNRAS.474.3808V}  (2) \citet{2012ApJ...758L..10M} (3) \citet{1993ApJ...416..208M}  \\
  }
  
  $^*$ Calculated in the frequency range of observations.
\end{tablenotes}
\end{threeparttable}
\end{center}
\end{table*} 

\subsection{IRAS~18162-2048}

This is the central region of the protostellar jet, and in radio wavelengths, the spectrum could include thermal free-free emission from the HII region ionized by stellar UV photons emitted by the YSO. Recently, it has been demonstrated that the HII region is likely to be in a nascent stage \citep{2020ApJ...888...41A}. We therefore proceed with the assumption that the free-free contribution to the spectrum of this source is solely due to the jet. \citet{2012ApJ...752L..29C} conducted high resolution observation of the region at a wavelength of 1.3~cm. We assume that the resolution of this observation ($0.1^{\prime \prime}$) corresponds to the projected distance of the base of the jet from the central source. The major axis of the central jet is $0.23^{\prime \prime}$, which we assume as the size (length) of the knot. It is possible to estimate the opening angle of the knot from its angular size (width) and radial distance using the formula, $\theta_0=2\tan^{-1}(\frac{\theta_{min}}{2 r_0})$~radians~$=38.5^\circ$, for $\theta_{min}$ = 0.07$^{\prime \prime}$. Here, $\theta_{min}$ represents the minor axis of the projected jet (perpendicular to the jet axis) as obtained from observations. We take this to be the width of the knot. Millimeter and sub-millimeter observations towards the central region of the jet have shown that the number density observed here is approximately in the range $10^{4} - 10^{5}$~cm$^{-3}$ \citep{2000AJ....119.2711H,2019ApJ...871..141Q,2019ApJ...877..112C}.

\begin{figure*}
\hspace*{0.5cm}
\minipage{\textwidth}
 \includegraphics[scale = 0.5]{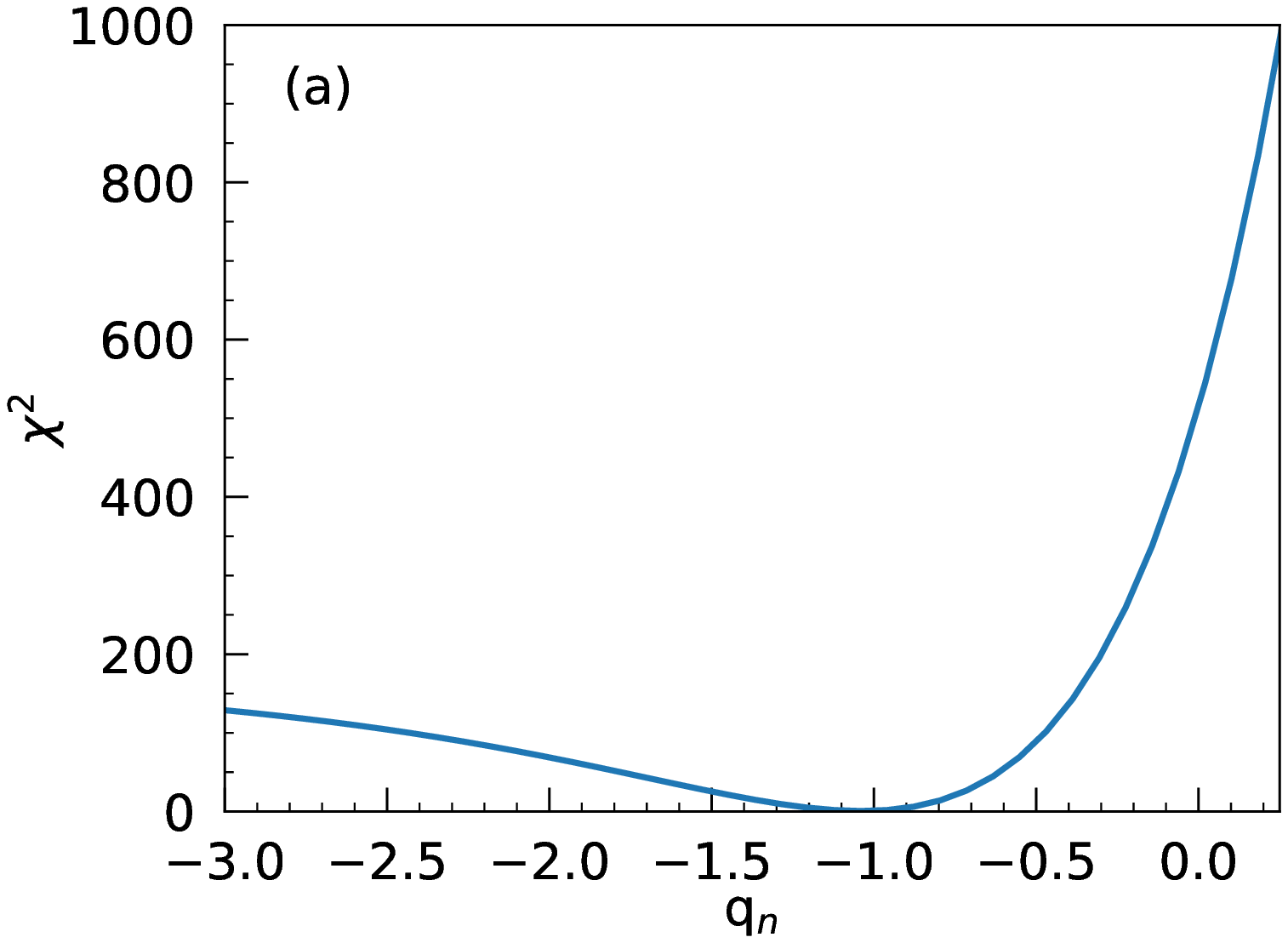}
\endminipage
\hspace*{-9.3cm}
\minipage{\textwidth}
  \includegraphics[scale = 0.5]{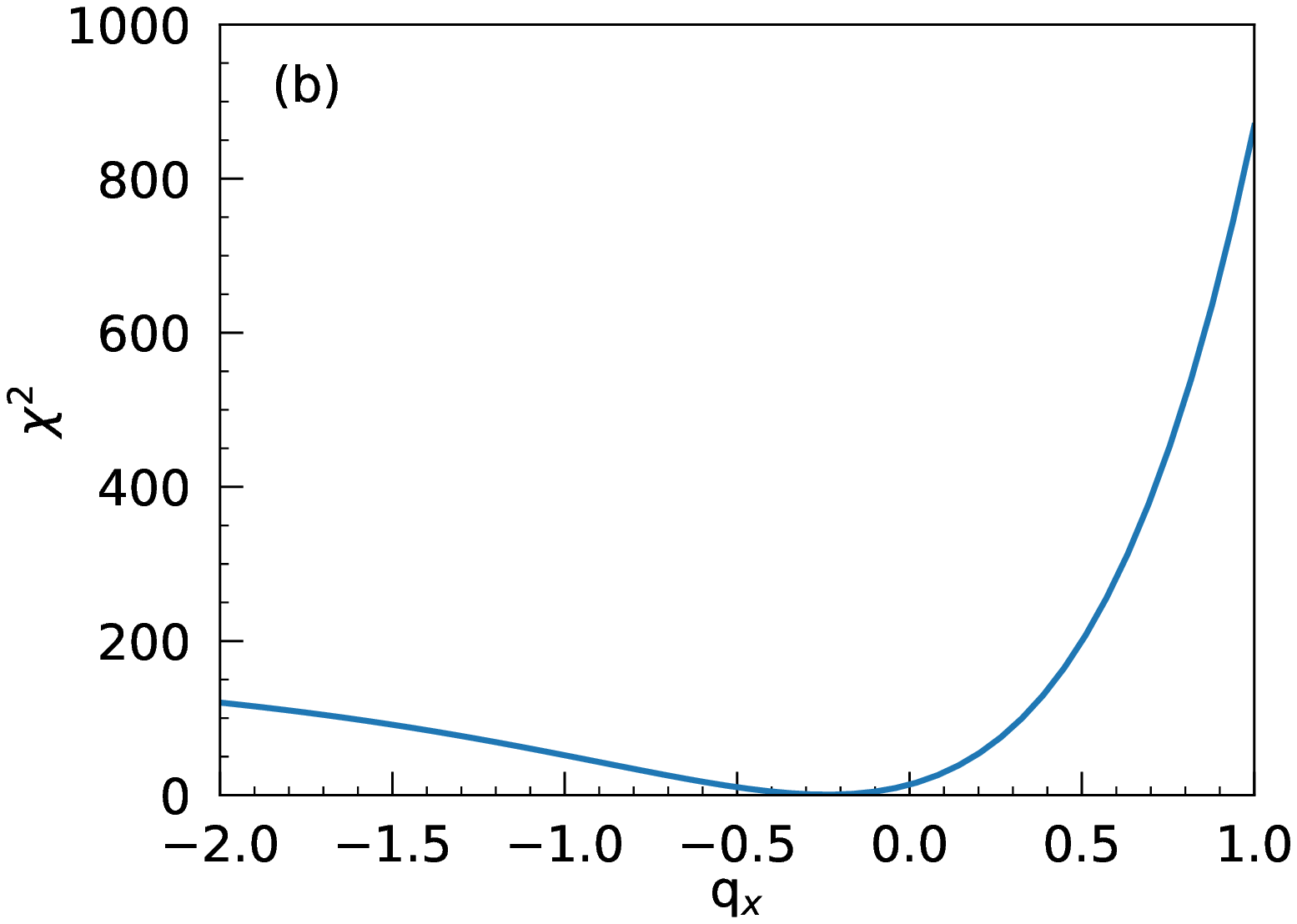}
\endminipage
\centering
\vskip +0.1cm
  \hspace*{0.5cm}
\minipage{\textwidth}
  \includegraphics[scale = 0.5]{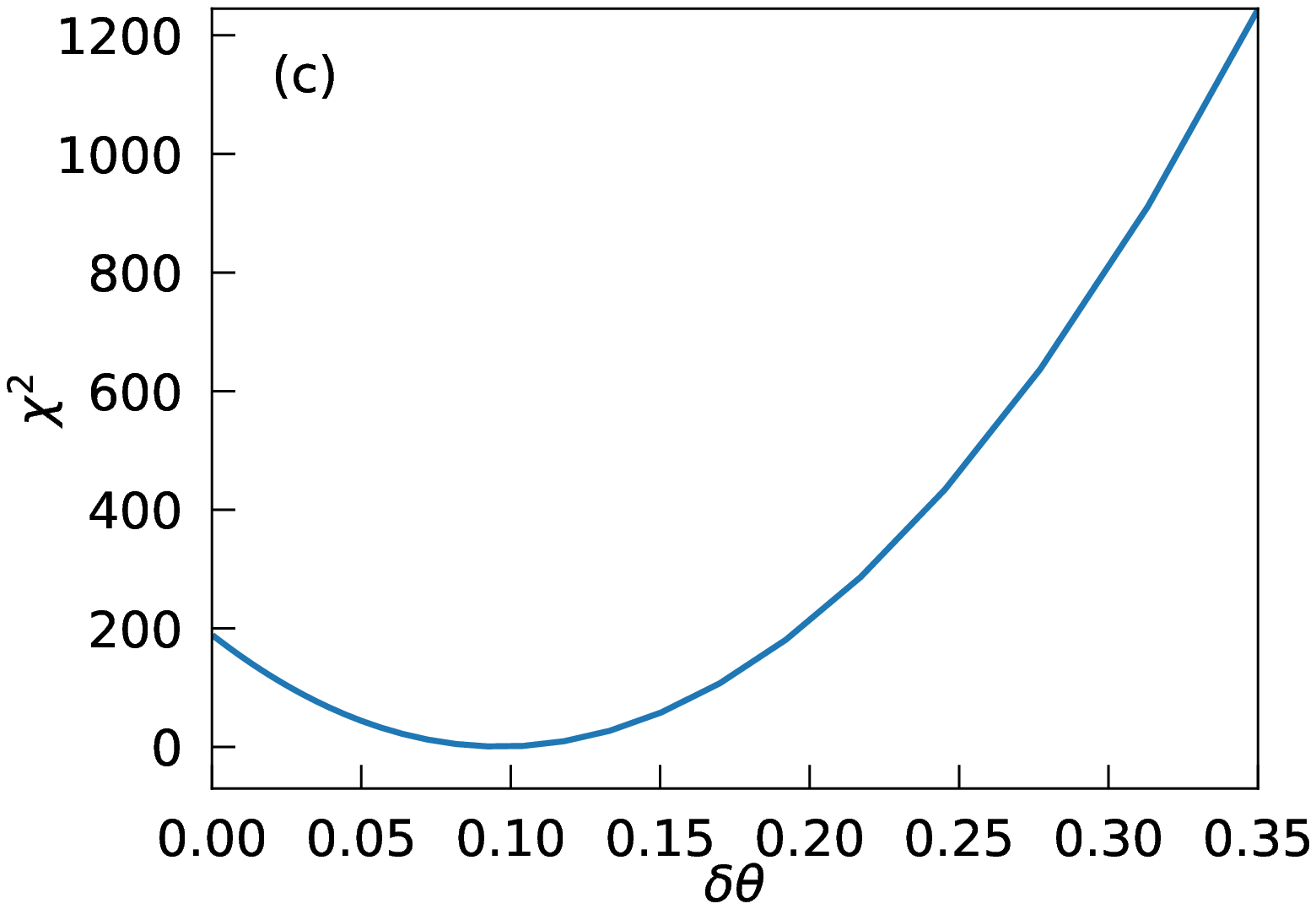}
\endminipage
\hspace*{-9.1cm}
\minipage{\textwidth}
  \includegraphics[scale = 0.5]{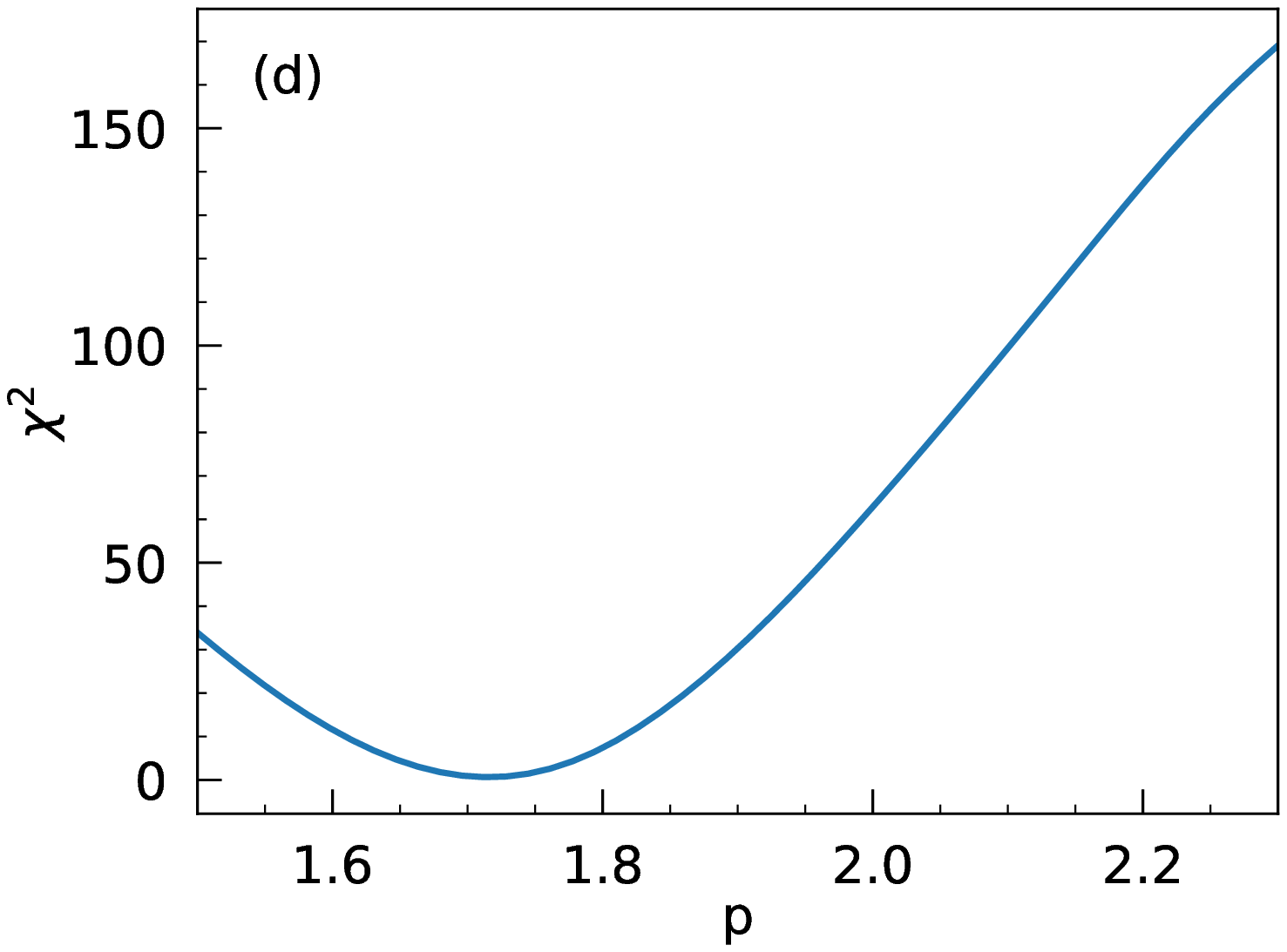}
\endminipage
\centering
\vskip +0.1cm
  \hspace*{0.4cm}
\minipage{\textwidth}
  \includegraphics[scale = 0.5]{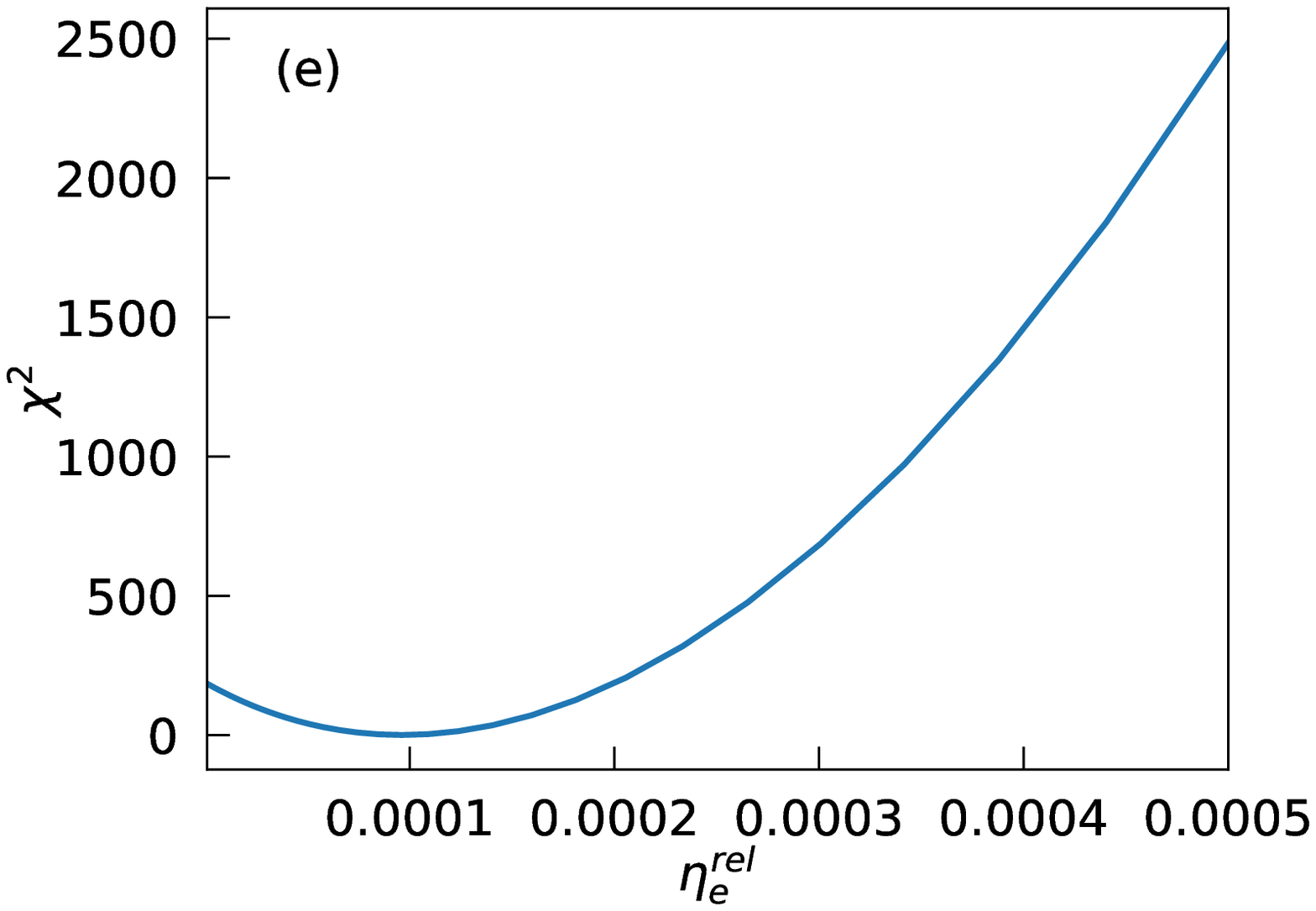}
\endminipage
\hspace*{-9.35cm}

\caption{$\chi^2$ value plots of the model 18162/2016 for parameters (a) q$_n$, (b) q$_x$, (c) $\delta \theta$, (d) $p$, and (e) $\eta^{rel}_{e}$, calculated by varying each parameter while the other best-fitting parameters are held constant. }
\label{fig:chi_sq1}
\end{figure*}

We first model the observations for data from the most recent epoch (2016) at low frequencies from \citet{2018MNRAS.474.3808V}. The best-fit model, designated as 18162/2016, is shown in Fig.~\ref{fig:model_spec}~(a) as the solid black curve, along with the observational data. This model reproduces the flat spectral index observed at low radio frequencies, which is explicable by a combination of thermal and non-thermal emission. The spectrum peaks at a frequency of $762$~MHz with a flux density of $2.75$~mJy. The spectral index of the model between $0.3 - 1.3$~GHz is $+0.13$. In order to demonstrate the fidelity of the determination method of the best-fitting parameters, we display the $\chi^2$ values of the parameter space that we have explored in the fitting procedure in Fig. ~\ref{fig:chi_sq1}. For this, we have varied each parameter while keeping the other parameters fixed to their best-fit values. The parameters under consideration for this knot include q$_n$, q$_x$, $\delta \theta$, $p$ and $\eta^{rel}_{e}$.

We also model the higher frequency VLA data corresponding to the observational epoch of 2009 \citep{2012ApJ...758L..10M}. The best-fit model, designated as 18162/2009, is shown in Fig.~\ref{fig:model_spec}~(a) as dashed grey curve, along with the observational data. For this model, we have assumed the same thickness of shocked shell ($\delta \theta= 0.1^\circ$) as obtained for the low frequency case. This spectrum peaks at a frequency of $5.3$~GHz with a flux density of $4.18$~mJy. The spectral index of the model calculated between $1.5 - 5.0$~GHz is $+0.11$.

The best-fit parameters for the models 18162/2016 and 18162/2009 are listed in Table.~\ref{tab:table_modelParms_assumed}. It can be seen that the change in the values of best-fit parameters between the two models is minor. The best fit values for $\delta \theta$, $p$ and $\eta^{rel}_{e}$ are same, whereas, $q_n$ varies in the range $-1$ to $-2$ and $q_x$ varies from $-0.2$ to $-0.5$ between the two models. This suggests a minor change in the density and ionization fraction structure. The best fit value of $\delta \theta$ shows that the shocked region comprises 0.8$\%$ of the jet opening angle.

\subsection{HH80}

The HH80 knot is located at a projected distance of $4.4^{\prime}$ from the central source. High angular resolution observation by \citet{1998AJ....116.1940H} has identified a large bright knot HH80A in this region with a projected linear size to be $1.5\times 10^{17}$~cm~($0.05$~pc). We take this to be approximately the size of the whole HH80 region that we are considering here. For the knot, we calculate an opening angle of $2.4^\circ$ using the opening angle of the central knot IRAS~18162-2048. The electron density in HH80 region is determined to be $200$~cm$^{-3}$ based on the [SII] line ratios \citep{1998AJ....116.1940H}. Since there are multiple epochs of observations at different frequencies and the jet is variable, we estimate the parameters for models corresponding to two observational epochs separated by nearly three decades: 2016 \citep{2018MNRAS.474.3808V} and 1989 \citep{1993ApJ...416..208M}.

\begin{figure*}
\hspace*{1cm}
\minipage{\textwidth}
 \includegraphics[scale = 0.5]{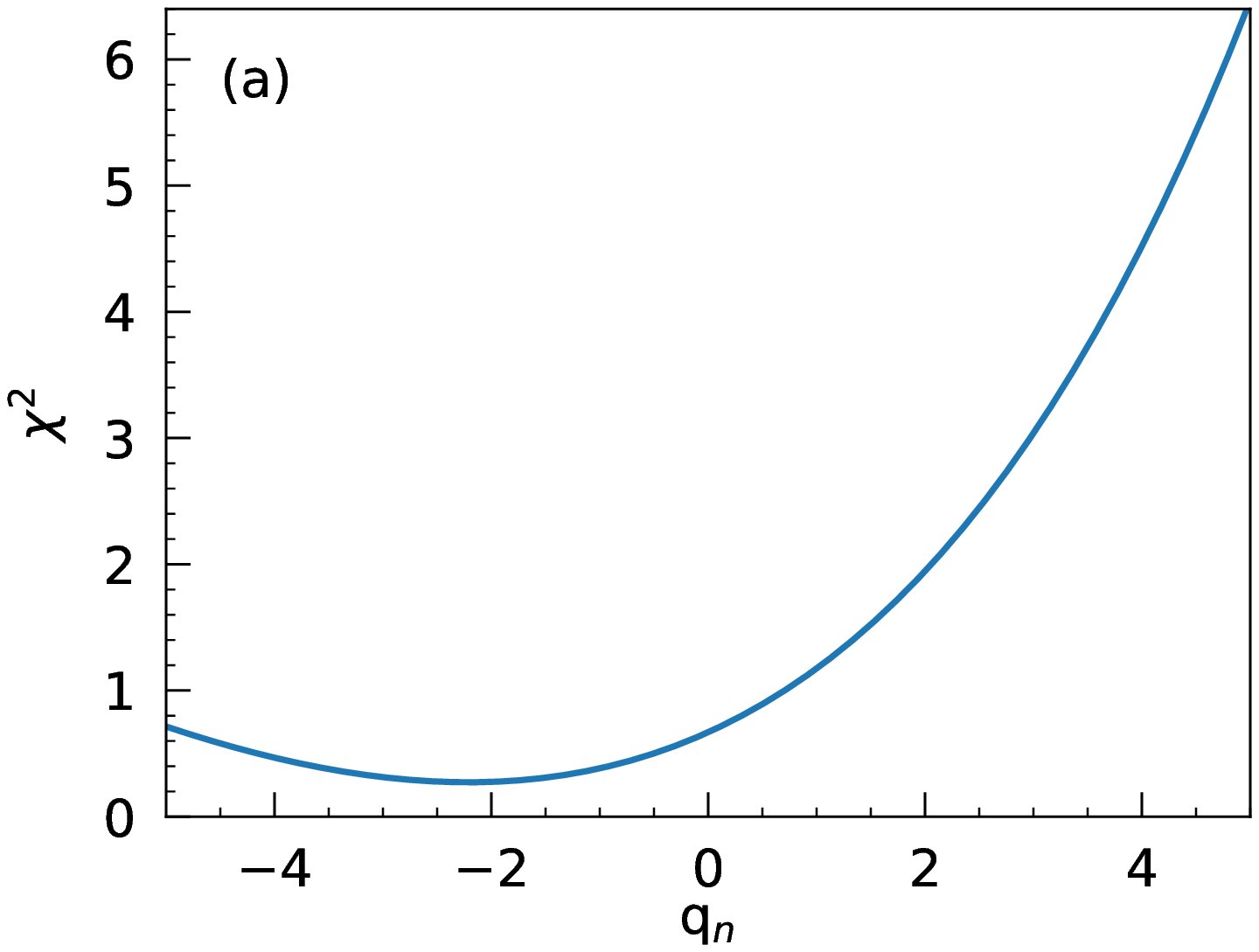}
\endminipage
\hspace*{-9.3cm}
\minipage{\textwidth}
  \includegraphics[scale = 0.5]{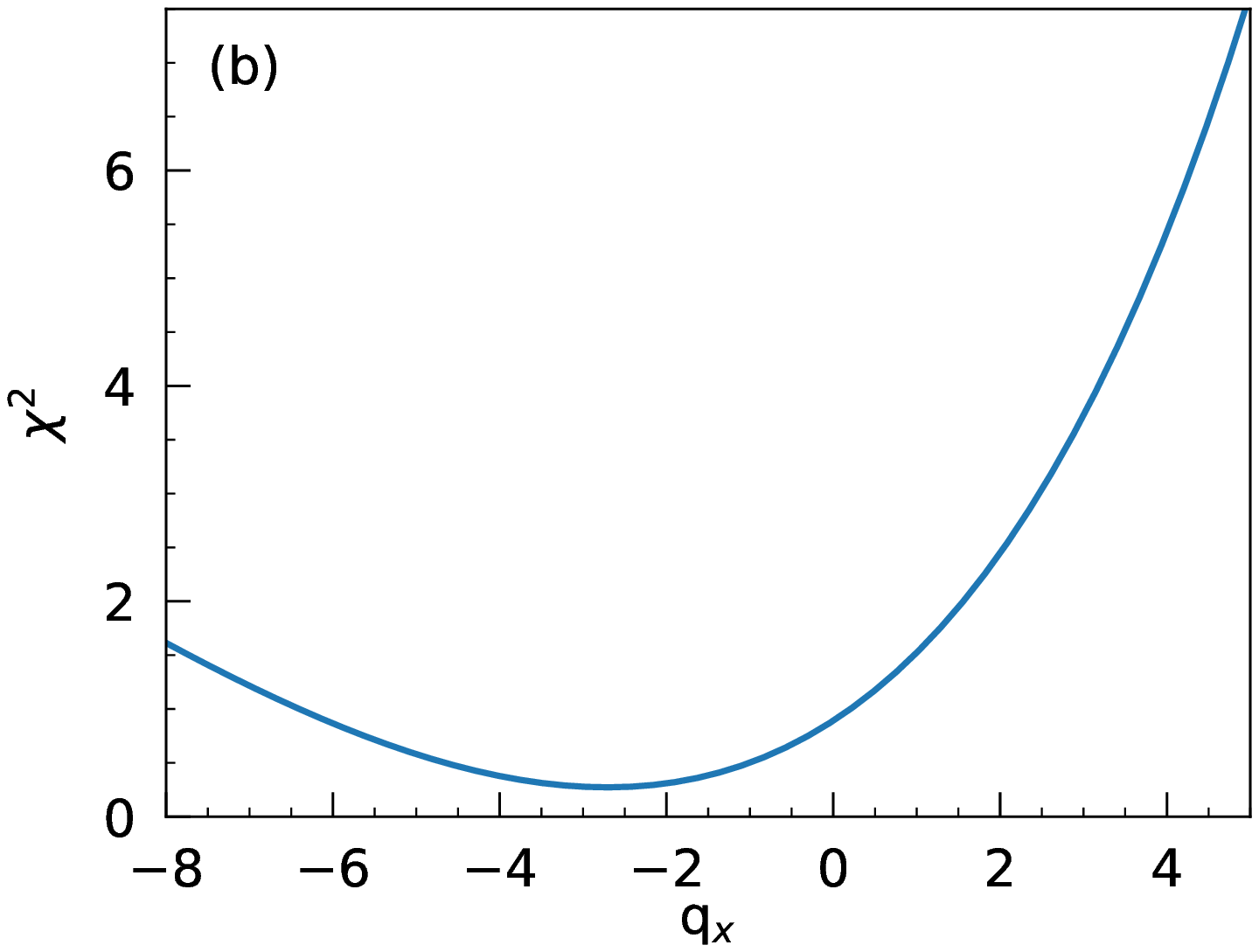}
\endminipage
\centering
\vskip +0.1cm
  \hspace*{1cm}
\minipage{\textwidth}
  \includegraphics[scale = 0.5]{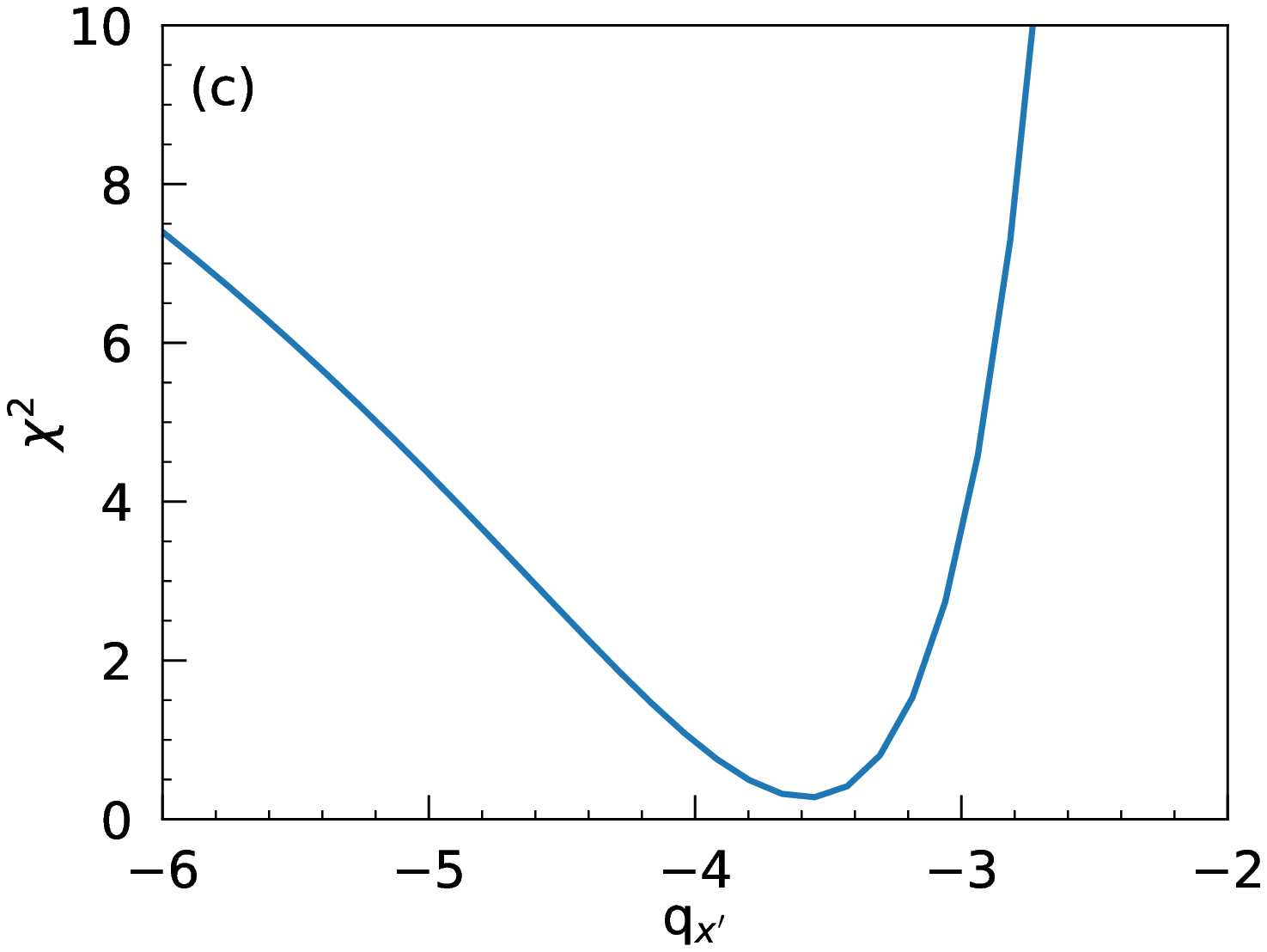}
\endminipage
\hspace*{-9.6cm}
\minipage{\textwidth}
  \includegraphics[scale = 0.5]{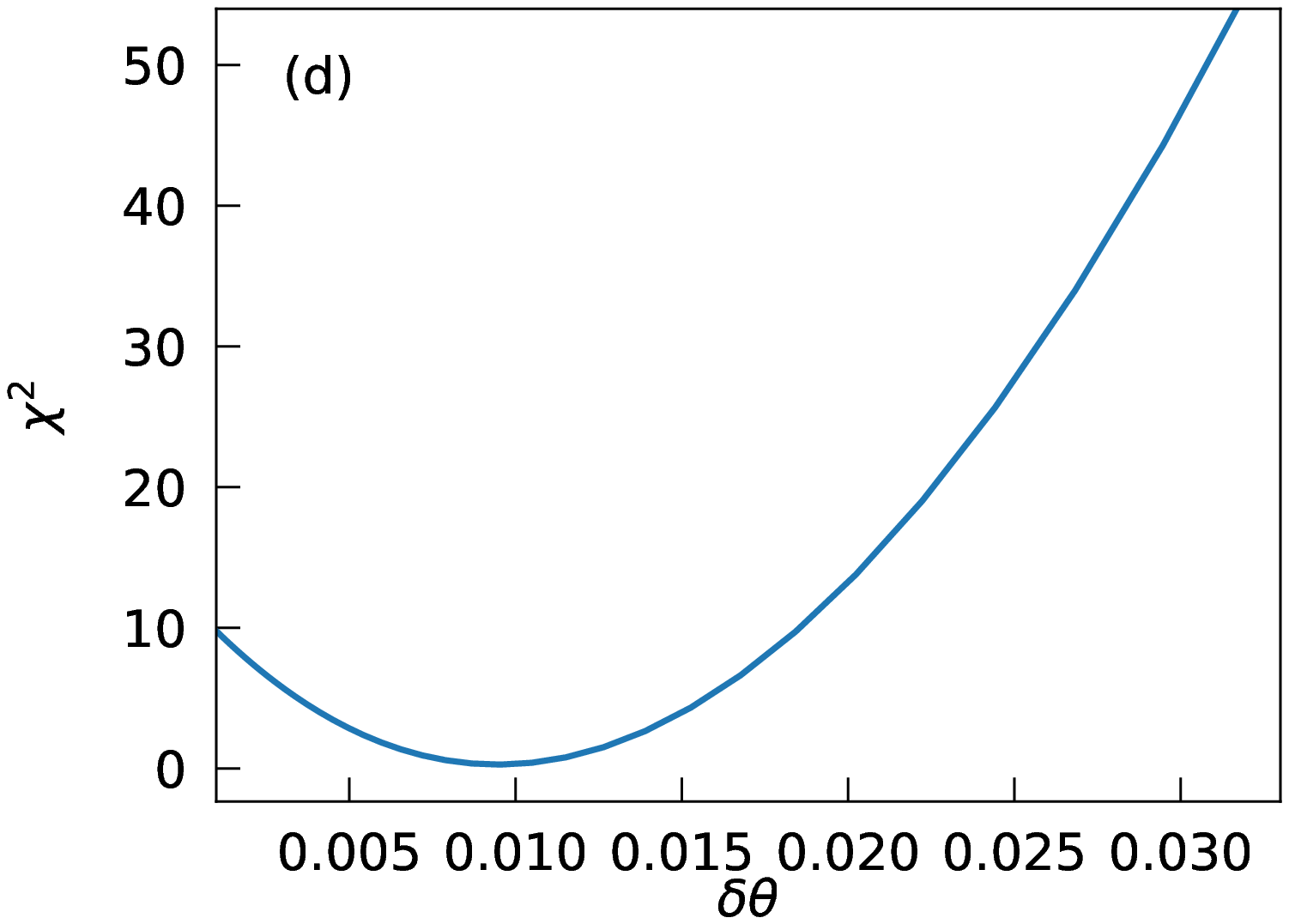}
\endminipage
\centering
\vskip +0.1cm
  \hspace*{0.8cm}
\minipage{\textwidth}
  \includegraphics[scale = 0.5]{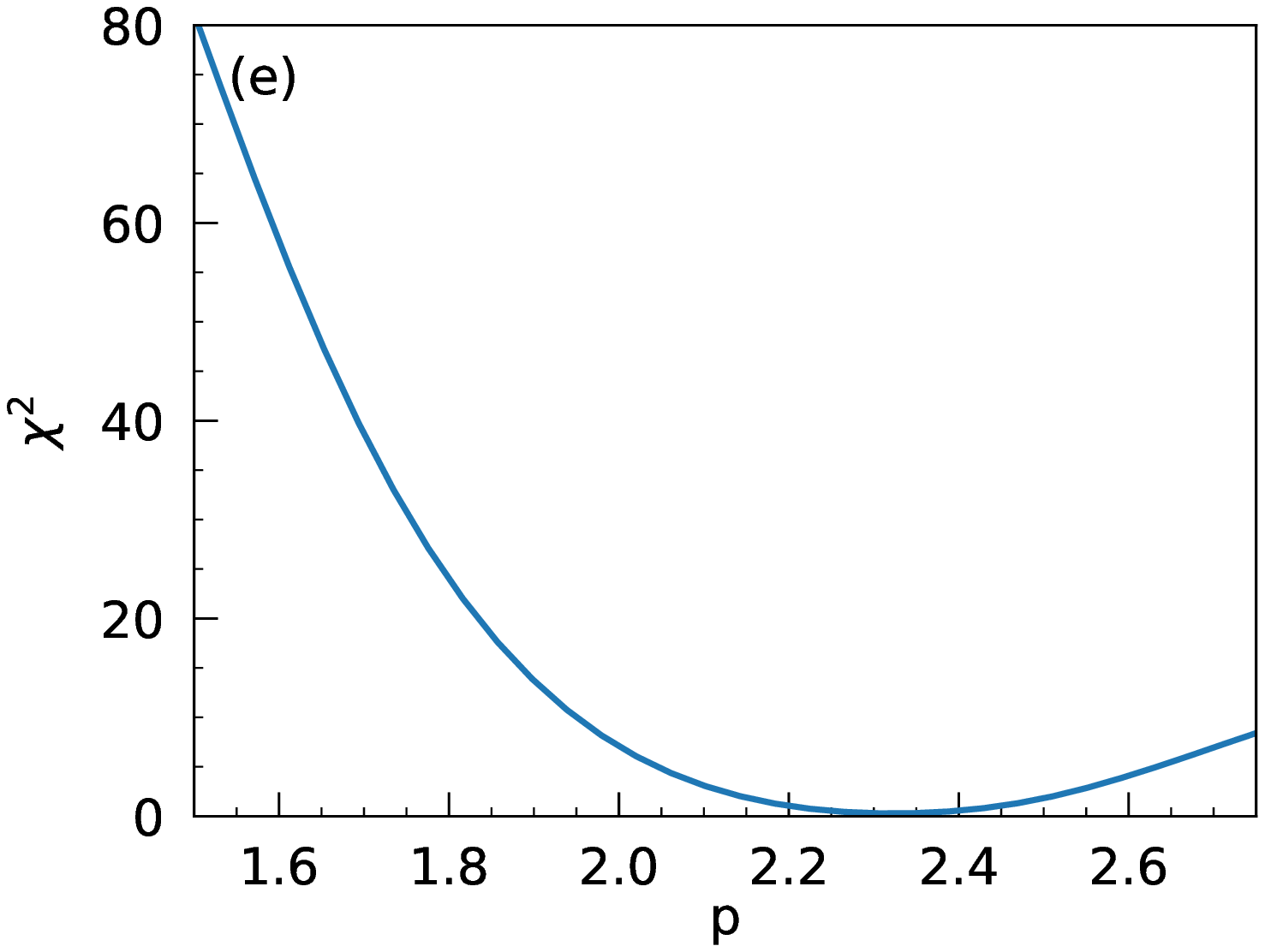}
\endminipage
\hspace*{-9.35cm}
\minipage{\textwidth}
  \includegraphics[scale = 0.5]{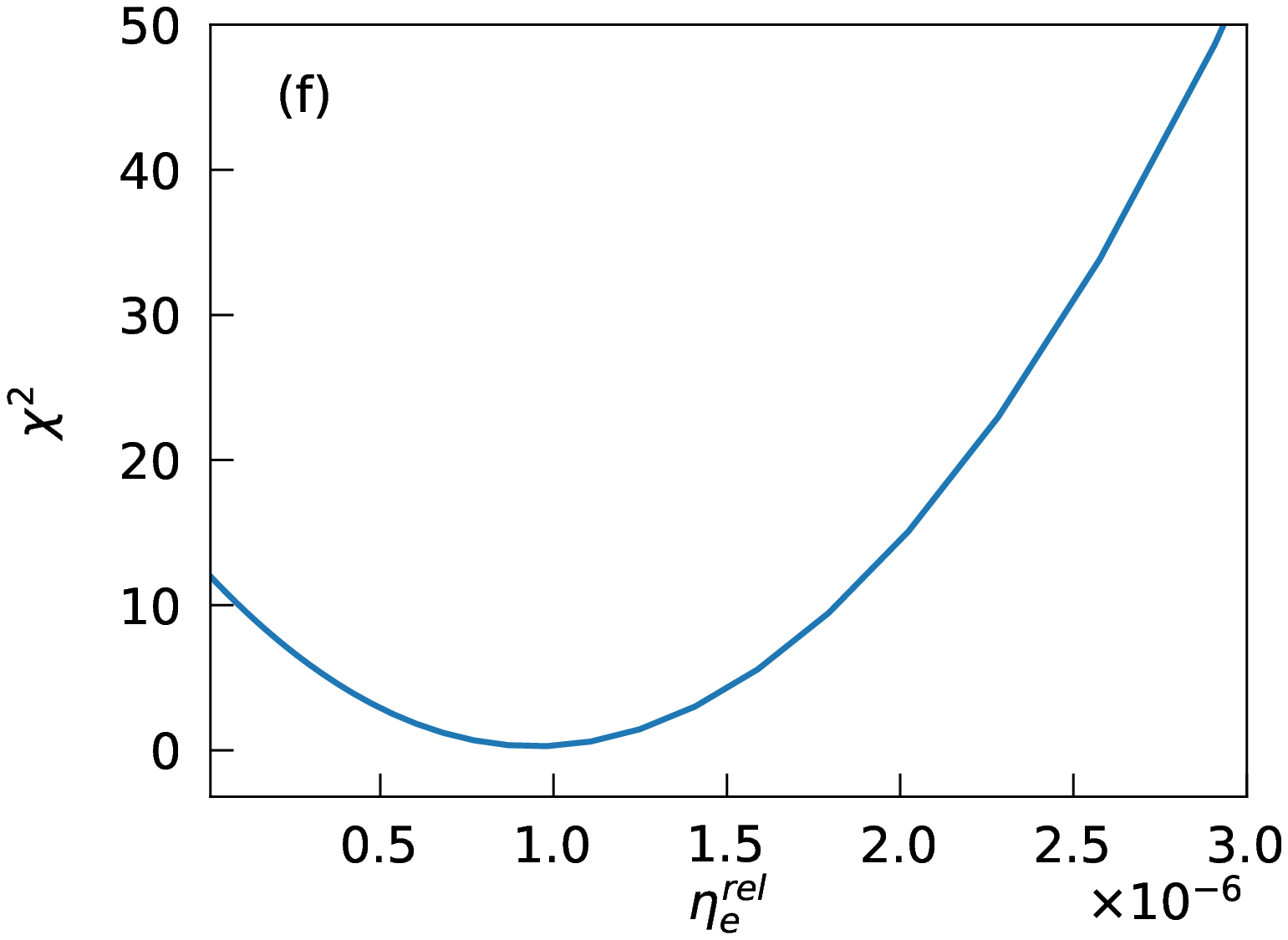}
\endminipage

\caption{$\chi^2$ value plots of the model HH80/2016 for parameters (a) q$_n$, (b) q$_x$, (c) q$_x^{\prime}$, (d) $\delta \theta$, (e) $p$, and (f) $\eta^{rel}_{e}$, calculated by varying each parameter while the other best-fitting parameters are held constant. }
\label{fig:chi_sq2}
\end{figure*}

The best-fit model for the low frequency observational data of \citet{2018MNRAS.474.3808V} designated as HH80/2016, is shown in Fig.~\ref{fig:model_spec}~(b) as the solid black curve, along with the data. The model spectrum peaks at a frequency of $171$~MHz with a flux density of $1.20$~mJy. The spectral index of the model between $0.3 - 1.3$~GHz is $-0.52$. The best-fit model provides a reasonably good fit with the observational data. The $\chi^2$ plots for the model parameters q$_n$, q$_x$, q$_x^{\prime}$, $\delta \theta$, $p$ and $\eta^{rel}_{e}$ are shown in Fig. ~\ref{fig:chi_sq2}.

The best-fit model for the high frequency observational data of \citet{1993ApJ...416..208M} designated as HH80/1989, is shown in Fig.~\ref{fig:model_spec}~(b) as the dashed grey curve, along with the data. For this epoch, we have assumed the same thickness of shocked shell ($\delta\theta=0.01^\circ$) as obtained for the low frequency case. The model spectrum peaks at a frequency of $13$~MHz with a flux density of $6.97$~mJy. The spectral index of the model between $1.5 - 15.0$~GHz is $-0.32$. The observing frequencies are different in the two epochs leading to different spectral indices for the two models.

The flux densities measured across three decades apart show that the emission from this knot displays variability. We find a difference of $0.5-1.8$~mJy in flux densities between $300$~MHz to $30$~GHz for the models HH80/2016 and HH80/1989. Therefore we have been able to reproduce the variability in observations using models that have a minor difference in jet parameters. The best-fit parameters for the models are listed in Table.~\ref{tab:table_modelParms_assumed}. For the two models, we have used the same electron number densities for the fitting procedure and obtained the best-fit values for the parameters to be similar, except for $q_x$ and $p$. These two parameters vary significantly between the two models, and this accounted for the flux density variation between the two epochs. The best fit value of $\delta \theta$ shows that the shocked region comprises 0.5$\%$ of the jet opening angle.

\subsection{HH81}

The HH81 knot is located at a projected distance of $3.7^{\prime}$ from the central protostar. The projected size of the bright HH81 knot is taken as $0.7\times 10^{17}$~cm \citep{1998AJ....116.1940H}, with an opening angle of $2.5^\circ$. Similar to HH80, we find best-fit models for HH81 corresponding to two epochs of observations to account for variability. The electron density in HH81 knot is assumed to be $1000$~cm$^{-3}$ which is similar to the densities obtained from the [SII] line ratios observed towards this region \citep{1998AJ....116.1940H}. 

\begin{figure*}
\hspace*{0.2cm}
\minipage{\textwidth}
 \includegraphics[scale = 0.5]{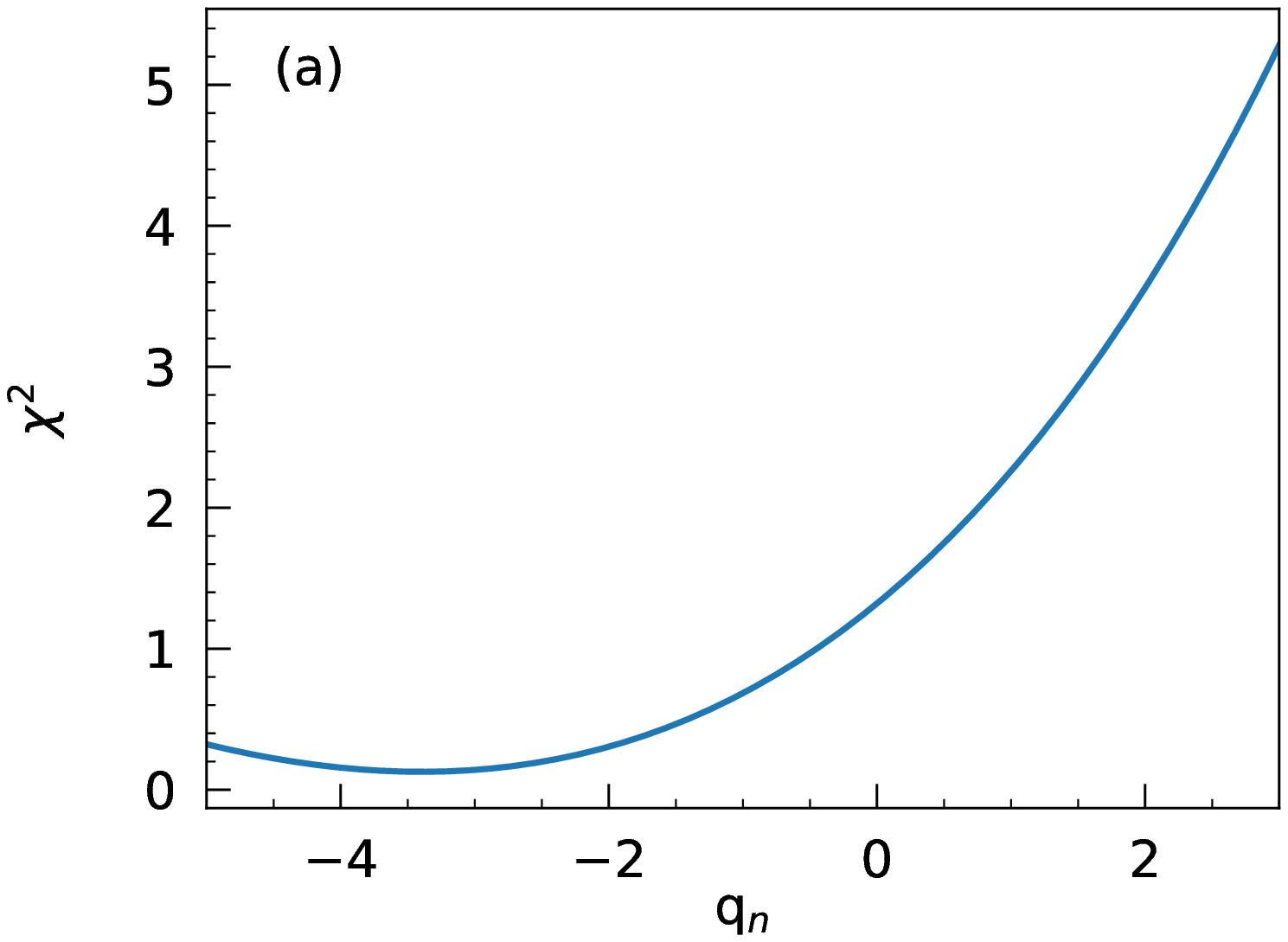}
\endminipage
\hspace*{-9.3cm}
\minipage{\textwidth}
  \includegraphics[scale = 0.5]{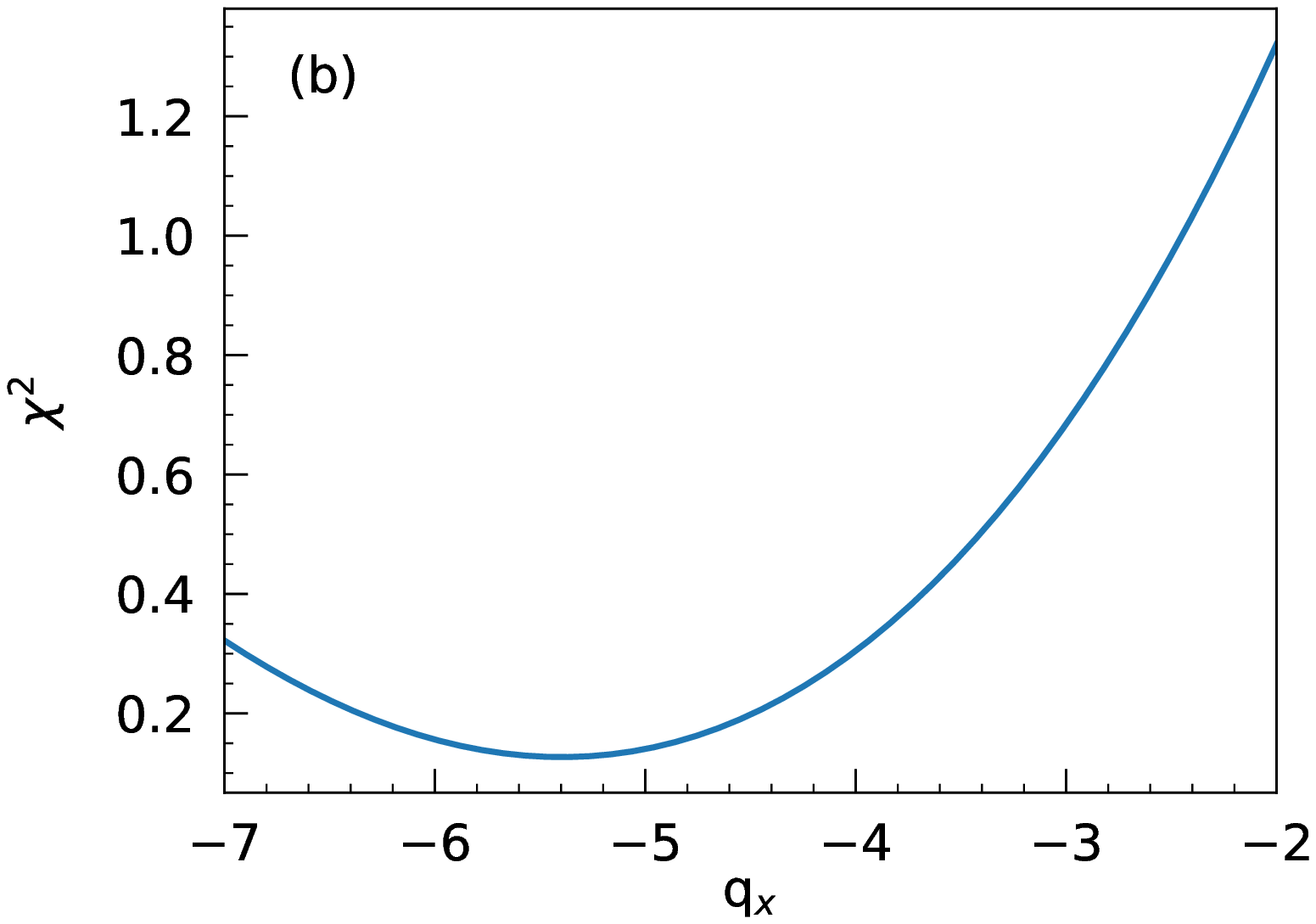}
\endminipage
\centering
\vskip +0.1cm
  \hspace*{0.5cm}
\minipage{\textwidth}
  \includegraphics[scale = 0.5]{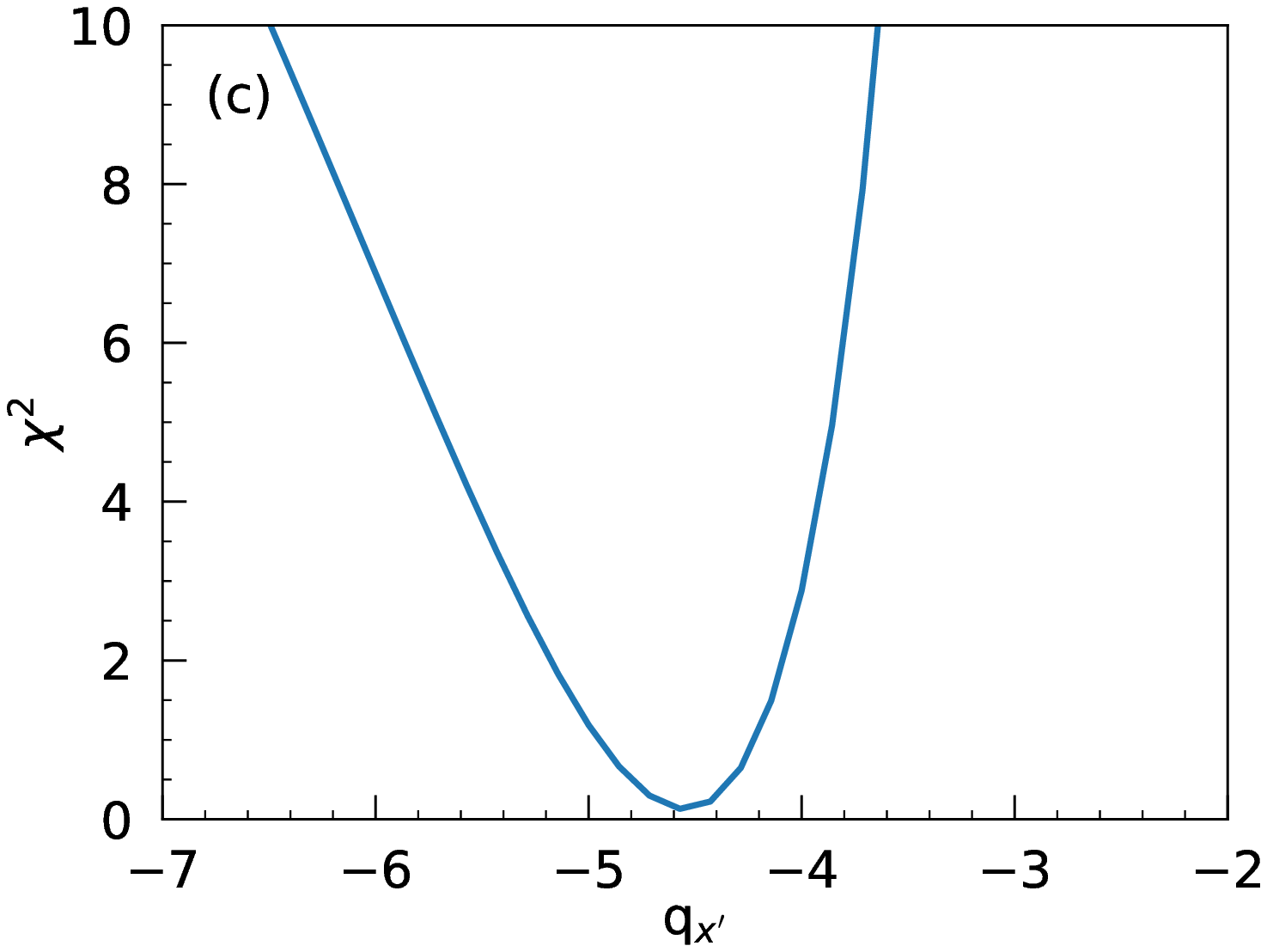}
\endminipage
\hspace*{-9.5cm}
\minipage{\textwidth}
  \includegraphics[scale = 0.5]{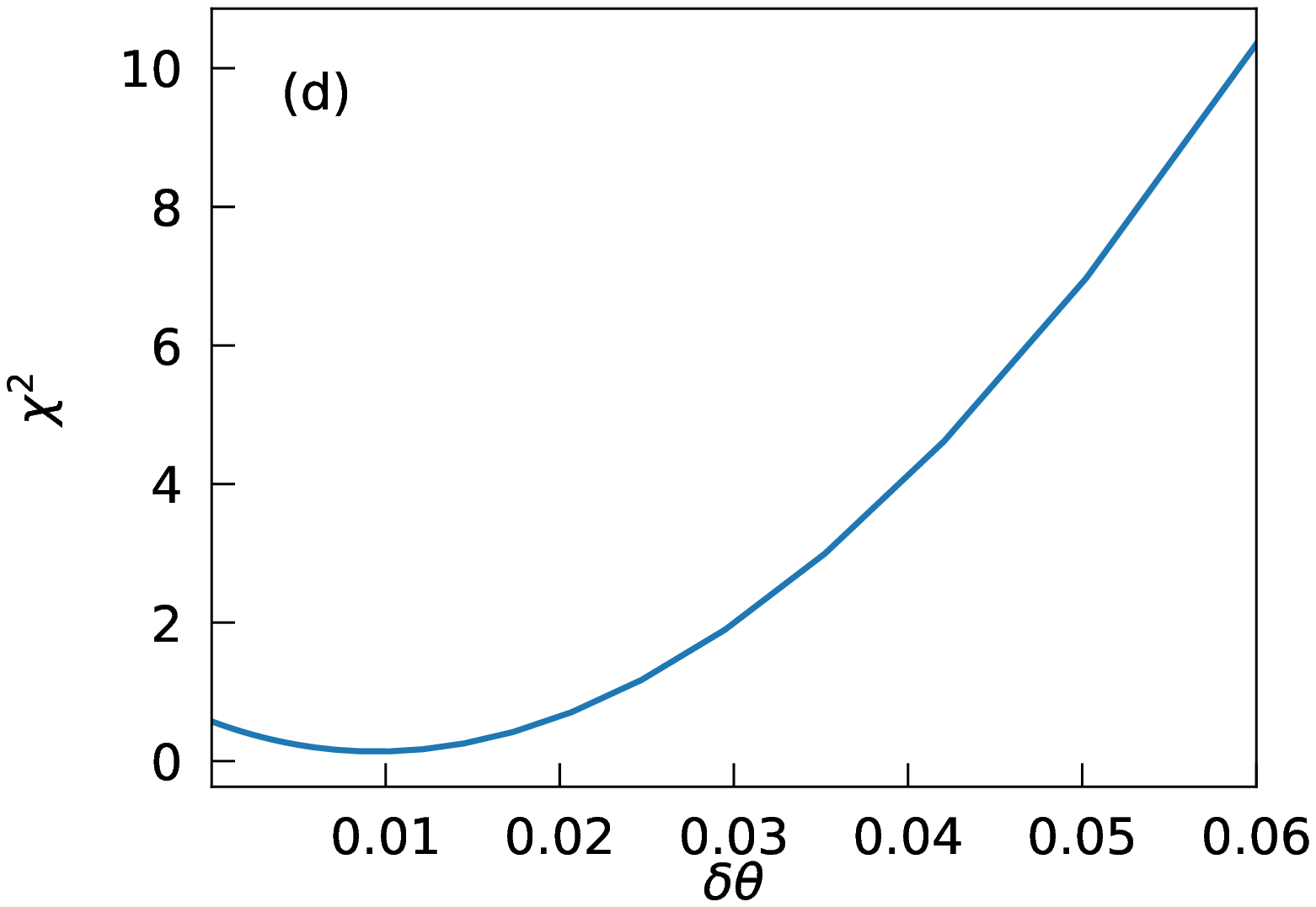}
\endminipage
\centering
\vskip +0.1cm
  \hspace*{0.4cm}
\minipage{\textwidth}
  \includegraphics[scale = 0.5]{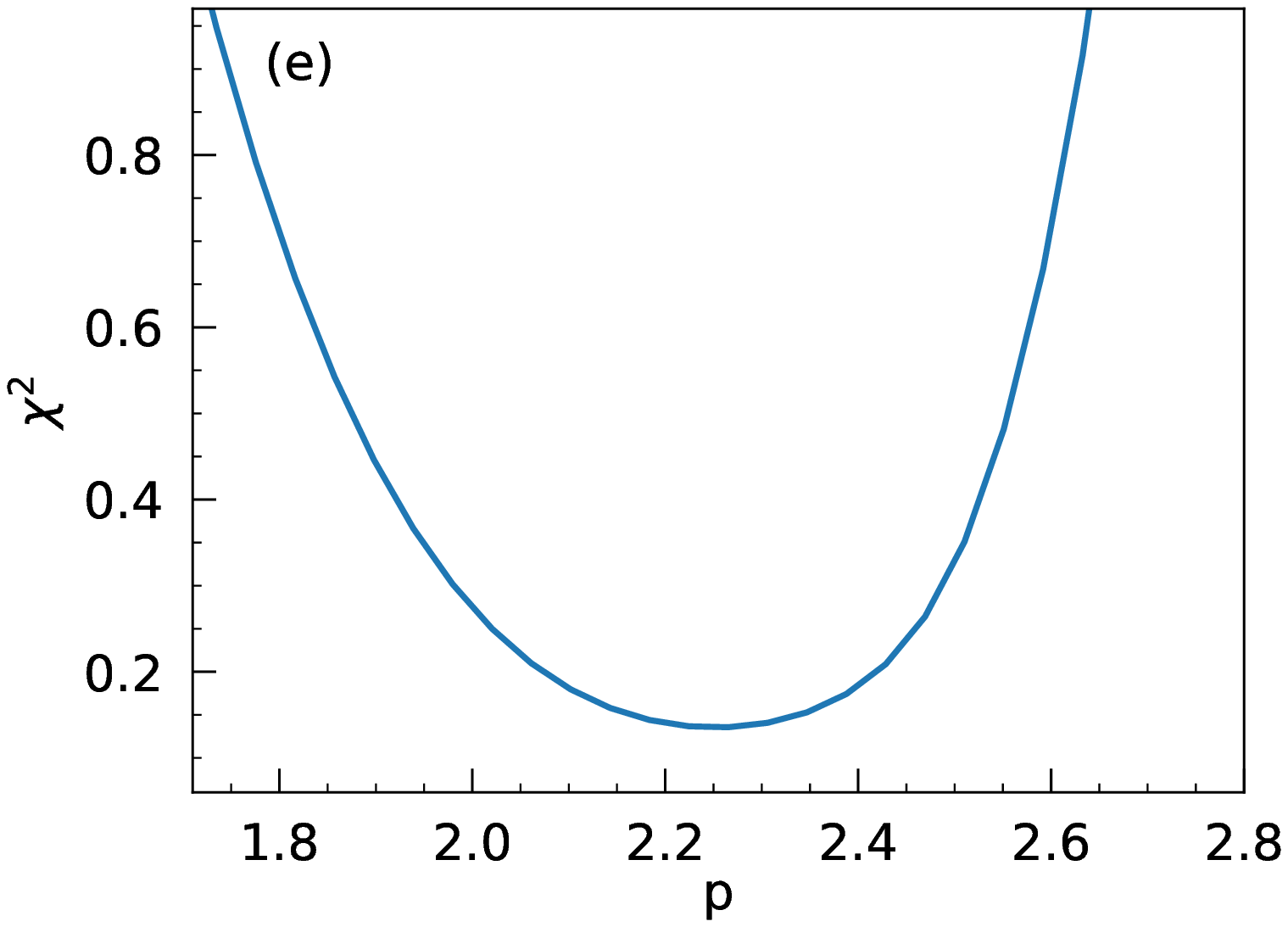}
\endminipage
\hspace*{-9.35cm}
\minipage{\textwidth}
  \includegraphics[scale = 0.5]{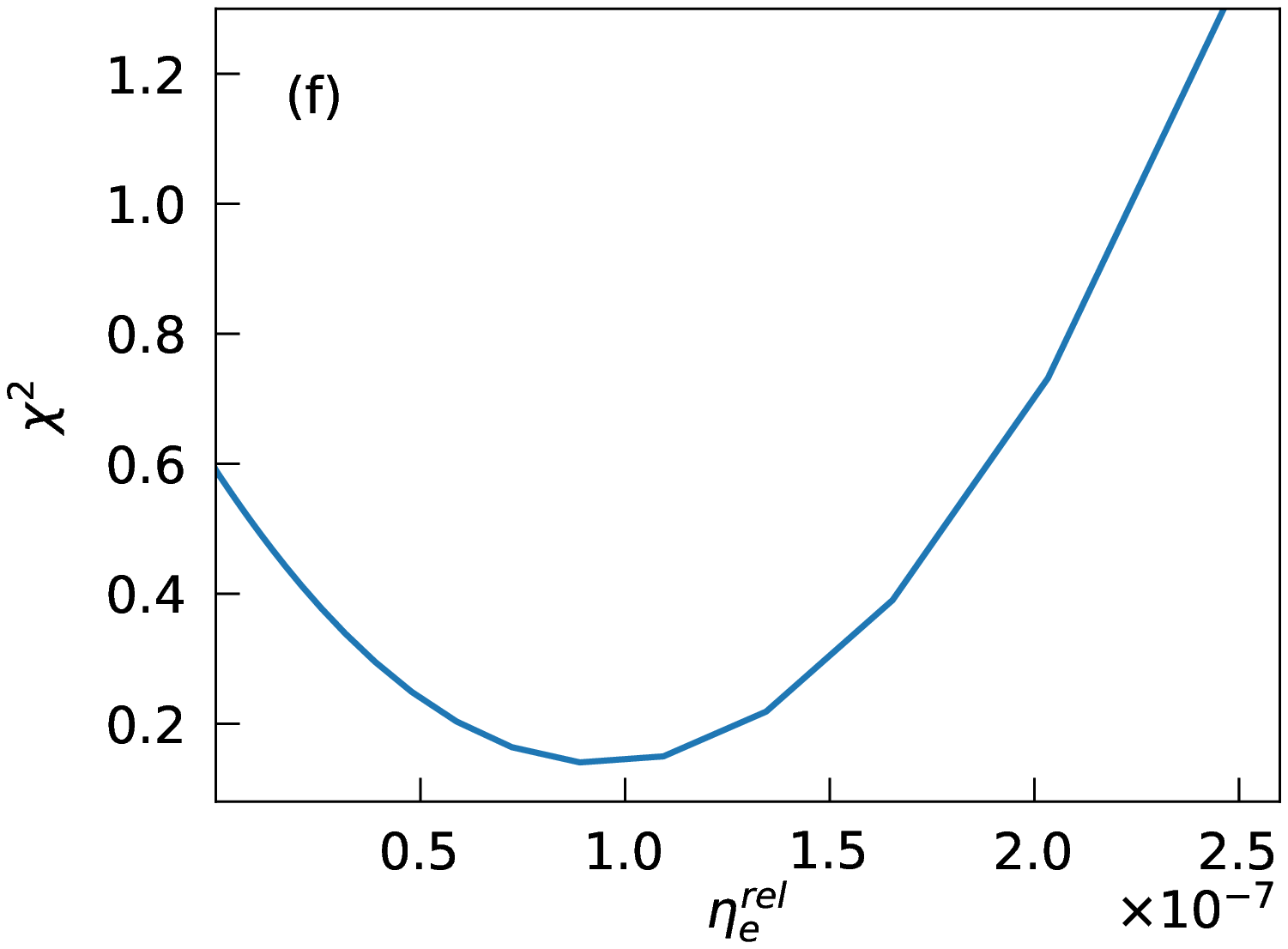}
\endminipage

\caption{$\chi^2$ value plots of the model HH81/2016 for parameters (a) q$_n$, (b) q$_x$, (c) q$_x^{\prime}$, (d) $\delta \theta$, (e) $p$, and (f) $\eta^{rel}_{e}$, calculated by varying each parameter while the other best-fitting parameters are held constant.}
\label{fig:chi_sq3}
\end{figure*}

The best-fit model for the low frequency observational data of \citet{2018MNRAS.474.3808V} designated as HH81/2016, is shown in Fig.~\ref{fig:model_spec}~(c) as the solid black curve, along with the data. The model spectrum peaks at a frequency of $147$~MHz with a flux density of $1.36$~mJy. The spectral index of the model between $0.3 - 1.3$~GHz is $-0.15$. The $\chi^2$ plots for the model parameters q$_n$, q$_x$, q$_x^{\prime}$, $\delta \theta$, $p$ and $\eta^{rel}_{e}$ are shown in Fig. ~\ref{fig:chi_sq3}.

The best-fit model for the high frequency observational data of \citet{1993ApJ...416..208M} designated as HH81/1989, is shown in Fig.~\ref{fig:model_spec}~(c) as the dashed grey curve, along with the data. For this epoch, we have assumed the same thickness of shocked shell ($\delta\theta=0.01^\circ$) as obtained for the low frequency case. The model spectrum peaks at a frequency of $486$~MHz with a flux density of $3.10$~mJy. The spectral index of the model between $1.5 - 15.0$~GHz is $-0.24$. 

Similar to the case of HH80 knot, the HH81 knot also displays variability as evident from the flux densities measured across three decades apart. We find a difference of $0.5-1.9$~mJy in flux densities between $300$~MHz to $30$~GHz for the models HH81/2016 and HH81/1989. Using models with a minor difference in the jet parameters, we have been able to replicate the variability for this knot as well. The best-fit parameters for the models are listed in Table.~\ref{tab:table_modelParms_assumed}. For the two models, we have used the same electron number densities for the fitting procedure and obtained the best-fit values for the parameters to be similar, except for $q_x$ which varies in the range $-0.2$ to $-5$. This is accounted to the flux density variation between the two epochs. The best fit value of $\delta \theta$ shows that the shocked region comprises 0.5$\%$ of the jet opening angle.\\

To summarize, in this section, we obtained the best-fit values for the model parameters $q_{x}$, $q_x'$, $p$, $\eta^{rel}_{e}$, $\delta \theta$ and $q_n$. For the knots HH80 and HH81, the best-fit values of $q_{x}$ show a significant variation between the model for two epochs, implying that it is an important parameter in deciding the nature of the spectrum. The jet material is primarily ionized in the radial direction due to internal shocks resulting from variations in the flow velocity. Large values of $q_x$ ($\sim -5$), therefore imply that the degree of ionization is significantly lower at farther radial distances from the driving source indicating a reduction in the strength of internal shocks which are capable of ionizing the jet material. Likewise, $q_x'$ indicates the degree of ionization resulting from shocks on the lateral edges of the jet where it impacts the ambient medium. This is also seen to weaken with radial distance from the driving source as inferred from the large values of $q_x'$ ($-4.5$ to $-3$) for HH80 and HH81 knots, compared to IRAS~18162-2048.

The radio spectra of HH80 and HH81 knots are dominated by synchrotron emission. The model parameters that strongly determine the relative contribution of synchrotron emission in the overall radio spectrum are $\eta^{rel}_{e}$ and $p$, which are representative of the efficiency of shock in generating a relativistic electron population and the density distribution of the non-thermal electron population in energy space, respectively. The best-fit values of $\eta^{rel}_{e}$ for HH80 and HH81, are in the range $10^{-7} - 10^{-6}$ which are comparable to the values obtained from the results of simulation studies in the context of diffusive shock acceleration \citep{1999ApJ...526..385B, 2016A&A...590A...8P}. This provides sufficient evidence to reinforce the assertion that the shocks present in these knots are efficient in generating relativistic electron population. The best-fit values of $p$ are in the range $1.7 - 2.4$, which are typical of shocks that are capable of generating non-thermal electron population in YSO jets \citep{2021MNRAS.504.2405A}. Given that this parameter $p$ is the power-law index of the number density distribution of the relativistic electron population in energy space, $p \leq 2.4$ could imply that the majority of the particle kinetic energy is carried by those particles with energies significantly higher than the average. 

HH80 and HH81 knots further demonstrates that despite having steeply decreasing power-law profiles for ionization fraction in the lateral direction (q$_{x'}= -3$ to $-4.5$) and a thin shocked region ($\frac{\delta \theta}{\theta} \sim 0.5\%$), the shock acceleration mechanism is highly efficient at generating non-thermal emission that dominates the total radio emission from these knots. This is evident from the best-fit values of the knot spectral indices which are in the range $-0.15$ to $-0.52$ (Table.~\ref{tab:specfeature}). For the central knot of the jet, we also observe that although $\eta^{rel}_{e}$ is much higher than HH80 and HH81 knots, the small fraction of shocked region with respect to the total opening angle ($\frac{\delta \theta}{\theta} \sim 0.8\%$) and the high number density (n$_0$) towards this region together results in higher contribution of thermal emission in the frequency window that was considered in this study.

For all the knots, the best-fit values obtained for the power-law index of number density profiles are the range q$_n = -1$ to $-3$. Previous observational studies and models have predicted a density profile with q$_n$ in the range $-1$ to $-2$ for the circumstellar material observed around YSOs \citep{1999ApJ...513..350H,1999ApJ...522..991V}. However, a steeper profile for number density (q$_n = -3$), for HH80 and HH81 knots, is also reasonable due to the fact that these knots are located much farther away \citep[$\sim 2-3$~pc;][]{1993ApJ...416..208M} from the driving source where the ambient density will be significantly lower than the immediate neighborhood of the central YSO. In addition, it was also observed that the corresponding knot in the northern arm of the jet, which is at similar radial distances from the driving source, has moved out of the molecular cloud \citep{1993ApJ...416..208M} into a region of lower density. Given this, we believe that the q$_n$ value obtained for these farther knots are feasible.


\section{Conclusion}
Protostellar jets are extremely collimated and are ejected during the accretion phase of star formation. In numerous cases, these jets are observed in the form of knots. In general, a combination of thermal free-free and non-thermal synchrotron emission processes can account for the observed radio spectrum of these knots. A numerical model that we had developed earlier takes into consideration both of these emission mechanisms and can be employed to generate radio spectra of protostellar jets. We apply this model to the case study of HH80-81 jet, and for this, we have considered the emission towards the central region of the jet, HH80 and HH81 knots. For each of these knots, we have generated best-fit models for two epochs corresponding to two frequency windows. Through this, we obtained various physical parameters of the knots, which are otherwise difficult to be solely determined from observations. The parameters obtained from the fitting include $q_n$, $q_x$, $q_x'$, $\delta \theta$, $p$ and $\eta^{rel}_{e}$ for the jet. Using electron number densities of $10^{3} - 10^{5}$~cm$^{-3}$ for these knots, we find the absolute values of power-law indices for ionization fraction profile in the radial and lateral direction to be in the range $0 - 5$. The jet edges that contribute to synchrotron emission contain non-thermal electron population with $\eta^{rel}_{e}$ in the range $10^{-7} - 10^{-4}$, which is consistent with the typical values observed in jet shocks. For the best-fit parameter sets, the model spectral indices lie in the range -0.15 to +0.11 within the observed frequency windows.

\section*{Acknowledgements}

We thank Carrasco-Gonzalez and Adriana Rodriguez-Kamenetzky for kindly sharing the VLA radio data of HH80-81 jet.

\bibliography{RadioJetModel}

\begin{thebibliography}{}
\expandafter\ifx\csname natexlab\endcsname\relax\def\natexlab#1{#1}\fi

\bibitem[{{A{\~n}ez-L{\'o}pez} {$et~al$.}(2020){A{\~n}ez-L{\'o}pez}, {Osorio},
  {Busquet}, {Girart}, {Mac{\'\i}as}, {Carrasco-Gonz{\'a}lez}, {Curiel},
  {Estalella}, {Fern{\'a}ndez-L{\'o}pez}, {Galv{\'a}n-Madrid}, {Kwon}, \&
  {Torrelles}}]{2020ApJ...888...41A}
{A{\~n}ez-L{\'o}pez}, N., {Osorio}, M., {Busquet}, G., {$et~al$.} 2020, \apj,
  888, 41

\bibitem[{{Araudo} {$et~al$.}(2021){Araudo}, {Padovani}, \&
  {Marcowith}}]{2021MNRAS.504.2405A}
{Araudo}, A.~T., {Padovani}, M., \& {Marcowith}, A. 2021, \mnras, 504, 2405

\bibitem[{Bellan {$et~al$.}(2005)Bellan, You, \& Hsu}]{Bellan2005}
Bellan, P.~M., You, S., \& Hsu, S.~C. 2005, Simulating Astrophysical Jets in
  Laboratory Experiments, ed. G.~Kyrala (Dordrecht: Springer Netherlands),
  203--209

\bibitem[{{Berezhko} \& {Ellison}(1999)}]{1999ApJ...526..385B}
{Berezhko}, E.~G., \& {Ellison}, D.~C. 1999, \apj, 526, 385

\bibitem[{{Biermann} \& {Strittmatter}(1987)}]{1987ApJ...322..643B}
{Biermann}, P.~L., \& {Strittmatter}, P.~A. 1987, \apj, 322, 643

\bibitem[{Blandford {$et~al$.}(2019)Blandford, Meier, \&
  Readhead}]{doi:10.1146/annurev-astro-081817-051948}
Blandford, R., Meier, D., \& Readhead, A. 2019, Annual Review of Astronomy and
  Astrophysics, 57, 467

\bibitem[{{Blandford} \& {Payne}(1982)}]{1982MNRAS.199..883B}
{Blandford}, R.~D., \& {Payne}, D.~G. 1982, \mnras, 199, 883

\bibitem[{{Bonito} {$et~al$.}(2010){Bonito}, {Orlando}, {Peres},
  {Eisl{\"o}ffel}, {Miceli}, \& {Favata}}]{2010A&A...511A..42B}
{Bonito}, R., {Orlando}, S., {Peres}, G., {$et~al$.} 2010, \aap, 511, A42

\bibitem[{{Bosch-Ramon} {$et~al$.}(2010){Bosch-Ramon}, {Romero}, {Araudo}, \&
  {Paredes}}]{2010A&A...511A...8B}
{Bosch-Ramon}, V., {Romero}, G.~E., {Araudo}, A.~T., \& {Paredes}, J.~M. 2010,
  \aap, 511, A8

\bibitem[{{Carrasco-Gonz{\'a}lez} {$et~al$.}(2010){Carrasco-Gonz{\'a}lez},
  {Rodr{\'\i}guez}, {Anglada}, {Mart{\'\i}}, {Torrelles}, \&
  {Osorio}}]{2010Sci...330.1209C}
{Carrasco-Gonz{\'a}lez}, C., {Rodr{\'\i}guez}, L.~F., {Anglada}, G., {$et~al$.}
  2010, Science, 330, 1209

\bibitem[{{Carrasco-Gonz{\'a}lez} {$et~al$.}(2012){Carrasco-Gonz{\'a}lez},
  {Galv{\'a}n-Madrid}, {Anglada}, {Osorio}, {D'Alessio}, {Hofner},
  {Rodr{\'\i}guez}, {Linz}, \& {Araya}}]{2012ApJ...752L..29C}
{Carrasco-Gonz{\'a}lez}, C., {Galv{\'a}n-Madrid}, R., {Anglada}, G., {$et~al$.}
  2012, \apjl, 752, L29

\bibitem[{Cerqueira \& de~Gouveia Dal~Pino(2001)}]{Cerqueira_2001}
Cerqueira, A.~H., \& de~Gouveia Dal~Pino, E.~M. 2001, The Astrophysical
  Journal, 550, L91

\bibitem[{{Cheng} {$et~al$.}(2019){Cheng}, {Qiu}, {Zhang}, {Wyrowski},
  {Menten}, \& {G{\"u}sten}}]{2019ApJ...877..112C}
{Cheng}, Y., {Qiu}, K., {Zhang}, Q., {$et~al$.} 2019, \apj, 877, 112

\bibitem[{{Dhawan} {$et~al$.}(2000){Dhawan}, {Mirabel}, \&
  {Rodr{\'\i}guez}}]{2000ApJ...543..373D}
{Dhawan}, V., {Mirabel}, I.~F., \& {Rodr{\'\i}guez}, L.~F. 2000, \apj, 543, 373

\bibitem[{{Felli} {$et~al$.}(2006){Felli}, {Massi}, {Robberto}, \&
  {Cesaroni}}]{2006A&A...453..911F}
{Felli}, M., {Massi}, F., {Robberto}, M., \& {Cesaroni}, R. 2006, \aap, 453,
  911

\bibitem[{{Girart} {$et~al$.}(2001){Girart}, {Estalella}, {Viti}, {Williams},
  \& {Ho}}]{2001ApJ...562L..91G}
{Girart}, J.~M., {Estalella}, R., {Viti}, S., {Williams}, D.~A., \& {Ho},
  P.~T.~P. 2001, \apjl, 562, L91

\bibitem[{{Harris} \& {Krawczynski}(2006)}]{2006ARA&A..44..463H}
{Harris}, D.~E., \& {Krawczynski}, H. 2006, \araa, 44, 463

\bibitem[{{Heathcote} {$et~al$.}(1998){Heathcote}, {Reipurth}, \&
  {Raga}}]{1998AJ....116.1940H}
{Heathcote}, S., {Reipurth}, B., \& {Raga}, A.~C. 1998, \aj, 116, 1940

\bibitem[{{Hogerheijde} {$et~al$.}(1999){Hogerheijde}, {van Dishoeck},
  {Salverda}, \& {Blake}}]{1999ApJ...513..350H}
{Hogerheijde}, M.~R., {van Dishoeck}, E.~F., {Salverda}, J.~M., \& {Blake},
  G.~A. 1999, \apj, 513, 350

\bibitem[{{Hunter} {$et~al$.}(2000){Hunter}, {Churchwell}, {Watson}, {Cox},
  {Benford}, \& {Roelfsema}}]{2000AJ....119.2711H}
{Hunter}, T.~R., {Churchwell}, E., {Watson}, C., {$et~al$.} 2000, \aj, 119,
  2711

\bibitem[{{Jhan} \& {Lee}(2016)}]{2016ApJ...816...32J}
{Jhan}, K.-S., \& {Lee}, C.-F. 2016, \apj, 816, 32

\bibitem[{{Koenigl}(1986)}]{1986NYASA.470...88K}
{Koenigl}, A. 1986, Annals of the New York Academy of Sciences, 470, 88

\bibitem[{Lee {$et~al$.}(2007)Lee, Ho, Hirano, Beuther, Bourke, Shang, \&
  Zhang}]{Lee_2007}
Lee, C.-F., Ho, P. T.~P., Hirano, N., {$et~al$.} 2007, The Astrophysical
  Journal, 659, 499

\bibitem[{{Lee} \& {Sahai}(2004)}]{2004ApJ...606..483L}
{Lee}, C.-F., \& {Sahai}, R. 2004, \apj, 606, 483

\bibitem[{{Li} \& {Nakamura}(2006)}]{2006ApJ...640L.187L}
{Li}, Z.-Y., \& {Nakamura}, F. 2006, \apjl, 640, L187

\bibitem[{{Livio}(1997)}]{1997ASPC..121..845L}
{Livio}, M. 1997, Astronomical Society of the Pacific Conference Series, Vol.
  121, {The Formation Of Astrophysical Jets}, ed. D.~T. {Wickramasinghe}, G.~V.
  {Bicknell}, \& L.~{Ferrario}, 845

\bibitem[{{Livio}(2000)}]{2000AIPC..522..275L}
{Livio}, M. 2000, in American Institute of Physics Conference Series, Vol. 522,
  American Institute of Physics Conference Series, ed. S.~S. {Holt} \& W.~W.
  {Zhang}, 275--297

\bibitem[{{Lovelace} {$et~al$.}(1987){Lovelace}, {Wang}, \&
  {Sulkanen}}]{1987ApJ...315..504L}
{Lovelace}, R.~V.~E., {Wang}, J.~C.~L., \& {Sulkanen}, M.~E. 1987, \apj, 315,
  504

\bibitem[{{Marti} {$et~al$.}(1993){Marti}, {Rodriguez}, \&
  {Reipurth}}]{1993ApJ...416..208M}
{Marti}, J., {Rodriguez}, L.~F., \& {Reipurth}, B. 1993, \apj, 416, 208

\bibitem[{{Marti} {$et~al$.}(1995){Marti}, {Rodriguez}, \&
  {Reipurth}}]{1995ApJ...449..184M}
---. 1995, \apj, 449, 184

\bibitem[{{Masqu{\'e}} {$et~al$.}(2012){Masqu{\'e}}, {Girart}, {Estalella},
  {Rodr{\'\i}guez}, \& {Beltr{\'a}n}}]{2012ApJ...758L..10M}
{Masqu{\'e}}, J.~M., {Girart}, J.~M., {Estalella}, R., {Rodr{\'\i}guez}, L.~F.,
  \& {Beltr{\'a}n}, M.~T. 2012, \apjl, 758, L10

\bibitem[{Meier {$et~al$.}(2001)Meier, Koide, \& Uchida}]{Meier84}
Meier, D.~L., Koide, S., \& Uchida, Y. 2001, Science, 291, 84

\bibitem[{Mirabel \& Rodríguez(1999)}]{doi:10.1146/annurev.astro.37.1.409}
Mirabel, I.~F., \& Rodríguez, L.~F. 1999, Annual Review of Astronomy and
  Astrophysics, 37, 409

\bibitem[{{Mohan} {$et~al$.}(2022){Mohan}, {Vig}, \&
  {Mandal}}]{2022MNRAS.514.3709M}
{Mohan}, S., {Vig}, S., \& {Mandal}, S. 2022, \mnras, 514, 3709

\bibitem[{{Nakamura} \& {Li}(2007)}]{2007ApJ...662..395N}
{Nakamura}, F., \& {Li}, Z.-Y. 2007, \apj, 662, 395

\bibitem[{{Padovani} {$et~al$.}(2016){Padovani}, {Marcowith}, {Hennebelle}, \&
  {Ferri{\`e}re}}]{2016A&A...590A...8P}
{Padovani}, M., {Marcowith}, A., {Hennebelle}, P., \& {Ferri{\`e}re}, K. 2016,
  \aap, 590, A8

\bibitem[{Palau {$et~al$.}(2014)Palau, Zapata, Rodríguez, Bouy, Barrado,
  Morales-Calderón, Myers, Chapman, Juárez, \& Li}]{10.1093/mnras/stu1461}
Palau, A., Zapata, L.~A., Rodríguez, L.~F., {$et~al$.} 2014, Monthly Notices
  of the Royal Astronomical Society, 444, 833

\bibitem[{{Price} {$et~al$.}(2003){Price}, {Pringle}, \&
  {King}}]{2003MNRAS.339.1223P}
{Price}, D.~J., {Pringle}, J.~E., \& {King}, A.~R. 2003, \mnras, 339, 1223

\bibitem[{{Pudritz} \& {Norman}(1983)}]{1983ApJ...274..677P}
{Pudritz}, R.~E., \& {Norman}, C.~A. 1983, \apj, 274, 677

\bibitem[{{Qiu} {$et~al$.}(2019){Qiu}, {Wyrowski}, {Menten}, {Zhang}, \&
  {G{\"u}sten}}]{2019ApJ...871..141Q}
{Qiu}, K., {Wyrowski}, F., {Menten}, K., {Zhang}, Q., \& {G{\"u}sten}, R. 2019,
  \apj, 871, 141

\bibitem[{Reipurth {$et~al$.}(2004)Reipurth, Rodrguez, Anglada, \&
  Bally}]{Reipurth_2004}
Reipurth, B., Rodrguez, L.~F., Anglada, G., \& Bally, J. 2004, The Astronomical
  Journal, 127, 1736

\bibitem[{{Reynolds}(1982)}]{1982ApJ...256...13R}
{Reynolds}, S.~P. 1982, \apj, 256, 13

\bibitem[{{Reynolds}(1986)}]{1986ApJ...304..713R}
---. 1986, \apj, 304, 713

\bibitem[{{Rodriguez} {$et~al$.}(1989){Rodriguez}, {Curiel}, {Moran},
  {Mirabel}, {Roth}, \& {Garay}}]{1989ApJ...346L..85R}
{Rodriguez}, L.~F., {Curiel}, S., {Moran}, J.~M., {$et~al$.} 1989, \apjl, 346,
  L85

\bibitem[{{Rodr{\'\i}guez} \& {Reipurth}(1989)}]{1989RMxAA..17...59R}
{Rodr{\'\i}guez}, L.~F., \& {Reipurth}, B. 1989, \rmxaa, 17, 59

\bibitem[{{Rodr{\'\i}guez-Kamenetzky}
  {$et~al$.}(2017){Rodr{\'\i}guez-Kamenetzky}, {Carrasco-Gonz{\'a}lez},
  {Araudo}, {Romero}, {Torrelles}, {Rodr{\'\i}guez}, {Anglada}, {Mart{\'\i}},
  {Perucho}, \& {Valotto}}]{2017ApJ...851...16R}
{Rodr{\'\i}guez-Kamenetzky}, A., {Carrasco-Gonz{\'a}lez}, C., {Araudo}, A.,
  {$et~al$.} 2017, \apj, 851, 16

\bibitem[{Rybicki \& Lightman(2008)}]{rybicki2008radiative}
Rybicki, G.~B., \& Lightman, A.~P. 2008, Radiative processes in astrophysics
  (John Wiley \& Sons)

\bibitem[{{Sari} {$et~al$.}(1998){Sari}, {Piran}, \&
  {Narayan}}]{1998ApJ...497L..17S}
{Sari}, R., {Piran}, T., \& {Narayan}, R. 1998, \apjl, 497, L17

\bibitem[{{Shimajiri} {$et~al$.}(2008){Shimajiri}, {Takahashi}, {Takakuwa},
  {Saito}, \& {Kawabe}}]{2008ApJ...683..255S}
{Shimajiri}, Y., {Takahashi}, S., {Takakuwa}, S., {Saito}, M., \& {Kawabe}, R.
  2008, \apj, 683, 255

\bibitem[{{Ustyugova} {$et~al$.}(1995){Ustyugova}, {Koldoba}, {Romanova},
  {Chechetkin}, \& {Lovelace}}]{1995ApJ...439L..39U}
{Ustyugova}, G.~V., {Koldoba}, A.~V., {Romanova}, M.~M., {Chechetkin}, V.~M.,
  \& {Lovelace}, R.~V.~E. 1995, \apjl, 439, L39

\bibitem[{{van der Tak} {$et~al$.}(1999){van der Tak}, {van Dishoeck}, {Evans},
  {Bakker}, \& {Blake}}]{1999ApJ...522..991V}
{van der Tak}, F. F.~S., {van Dishoeck}, E.~F., {Evans}, Neal~J., I., {Bakker},
  E.~J., \& {Blake}, G.~A. 1999, \apj, 522, 991

\bibitem[{{Vig} {$et~al$.}(2018){Vig}, {Veena}, {Mandal}, {Tej}, \&
  {Ghosh}}]{2018MNRAS.474.3808V}
{Vig}, S., {Veena}, V.~S., {Mandal}, S., {Tej}, A., \& {Ghosh}, S.~K. 2018,
  \mnras, 474, 3808

\bibitem[{{Whelan} {$et~al$.}(2009){Whelan}, {Ray}, {Bacciotti}, {Rand ich}, \&
  {Natta}}]{2009ASSP...13..259W}
{Whelan}, E.~M., {Ray}, T., {Bacciotti}, F., {Rand ich}, S., \& {Natta}, A.
  2009, Astrophysics and Space Science Proceedings, 13, 259

\bibitem[{{Wijers} \& {Galama}(1999)}]{1999ApJ...523..177W}
{Wijers}, R.~A.~M.~J., \& {Galama}, T.~J. 1999, \apj, 523, 177

\bibitem[{Zanni {$et~al$.}(2004)Zanni, Ferrari, Massaglia, Bodo, \&
  Rossi}]{zanni2004mhd}
Zanni, C., Ferrari, A., Massaglia, S., Bodo, G., \& Rossi, P. 2004,
  Astrophysics and Space Science, 293, 99

\end{thebibliography}

\end{document}